\documentclass[prd,twocolumn,showpacs,preprintnumbers,nofootinbib,amsmath,amssymb,superscriptaddress,nob
alancelastpage]{revtex4}
\usepackage{graphicx, color}

\newcommand{\sfig}[2]{
\includegraphics[width=#2]{#1}
		}
\newcommand{\Sfig}[2]{
	\begin{figure}[thbp]
	\sfig{#1.eps}{\columnwidth}
	\caption{{\small #2}}
	\label{fig:#1}
	\end{figure}
}

\definecolor{Blue}{rgb}{0.25,.41,.88}

\def\cmm2{{\,\rm cm^{-2}}}
\def\cm2{{\,{\rm cm}^2}}
\def\cmm3{{\,{\rm cm}^{-3}}}
\def\gcmm3{{\,{\rm g\,cm^{-3}}}}

\def\fun#1#2{\lower3.6pt\vbox{\baselineskip0pt\lineskip.9pt
  \ialign{$\mathsurround=0pt#1\hfil##\hfil$\crcr#2\crcr\sim\crcr}}}

\def\be{\begin{equation}}
\def\ee{\end{equation}}
\def\bea{\begin{eqnarray}}
\def\eea{\end{eqnarray}}

\newcommand\cov{\boldsymbol{\mathcal{C}}}
\newcommand\fish{\boldsymbol{\mathcal{F}}}

\newcommand{\dd}{\mathrm{d}}
\newcommand{\Omegam}{\Omega_{\mathrm{M}}}

\newcommand{\omegam}{\omega_{\mathrm{M}}}
\newcommand{\omegab}{\omega_{\mathrm{B}}}

\newcommand{\wzero}{w_{0}}
\newcommand{\wa}{w_{\mathrm{a}}}

\newcommand{\rhomean}{\rho_{\mathrm{M}}}

\newcommand{\Rvir}{R_{200\mathrm{m}}}

\newcommand{\rhovir}{\rho_{\mathrm{vir}}}

\newcommand{\Pktom}[2]{P_{\kappa}^{\mathrm{#1}\mathrm{#2}}}
\newcommand{\Pkobs}[2]{\bar{P}_{\kappa}^{\mathrm{#1}\mathrm{#2}}}

\newcommand{\ellmax}{\ell_{\mathrm{max}}}
\newcommand{\wlw}[1]{W_{\mathrm{#1}}}
\newcommand{\Da}{D_{\mathrm{A}}}

\newcommand{\ntomo}{N_{\mathrm{z}}}
\newcommand{\nobs}{N_{\mathrm{obs}}}

\newcommand{\fsky}{f_{\mathrm{sky}}}
\newcommand{\gi}{\langle \gamma^2 \rangle}

\usepackage{graphicx}
\graphicspath{{converted_graphics/}}
\begin{document}

\title{Accounting for Baryons in Cosmological Constraints from Cosmic Shear} % Enter your title between curly braces
\author{Andrew R. Zentner}% Enter your name between curly braces
\affiliation{Department of Physics and Astronomy, 
University of Pittsburgh, Pittsburgh, PA 15260}
\affiliation{PITTsburgh Particle physics, Astrophysics, and Cosmology Center (PITT PACC), 
University of Pittsburgh, Pittsburgh, PA 15260}
\author{Elisabetta Semboloni}
\affiliation{Leiden Observatory, Leiden University, P.O. Box 9513, 2300 RA, The Netherlands}
\author{Scott Dodelson}
\affiliation{Kavli Institute for Cosmological Physics,  Enrico Fermi Institute, University of Chicago, Chicago, IL 60637}
\affiliation{Department of Astronomy \& Astrophysics, University of Chicago, Chicago, IL 60637}
\affiliation{Fermilab Center for Particle Astrophysics, 
Fermi National Accelerator Laboratory, Batavia, IL 60510-0500}
\author{Tim Eifler}
\affiliation{Department of Physics and Astronomy, University of Pennsylvania, Philadelphia, PA 19104}
\affiliation{Center for Cosmology and Astro-Particle Physics, The Ohio State University, 
191 W. Woodruff Ave., Columbus, OH 43210}
\author{Elisabeth Krause}
\affiliation{Department of Physics and Astronomy, University of Pennsylvania, Philadelphia, PA 19104}
\author{Andrew P. Hearin}
\affiliation{Department of Physics and Astronomy, 
University of Pittsburgh, Pittsburgh, PA 15260}
\affiliation{PITTsburgh Particle physics, Astrophysics, and Cosmology Center (PITT PACC), 
University of Pittsburgh, Pittsburgh, PA 15260}
\affiliation{Fermilab Center for Particle Astrophysics, 
Fermi National Accelerator Laboratory, Batavia, IL 60510-0500}
\date{\today}% Enter your date or \today between curly braces

%------------------------------------------------------------------------
\begin{abstract}
One of the most pernicious theoretical systematics
facing upcoming gravitational lensing surveys is the uncertainty 
introduced by the effects of baryons on the power spectrum of 
the convergence field. 
One method that has been proposed to account for these effects 
is to allow several additional parameters 
(that characterize dark matter halos) to vary and 
to fit lensing data to these halo parameters concurrently with 
the standard set of cosmological parameters. We test this method. 
In particular, we use this technique to model 
convergence power spectrum predictions from a set of cosmological simulations. 
We estimate biases in dark energy equation of state parameters 
that would be incurred if one were to fit the spectra predicted by the 
simulations either with no model for baryons, or with the 
proposed method. We show that neglecting baryonic effect leads to biases 
in dark energy parameters that are several times the statistical errors 
for a survey like the Dark Energy Survey. The proposed method to 
correct for baryonic effects renders the residual biases in dark energy equation 
of state parameters smaller than the statistical errors. These results 
suggest that this mitigation method may be applied to analyze convergence 
spectra from a survey like the Dark Energy Survey. 
For significantly larger surveys, such as will be carried out by 
the Large Synoptic Survey Telescope, the biases introduced by baryonic 
effects are much more significant. We show that this mitigation technique 
significantly reduces the biases for such larger surveys, 
but that a more effective mitigation strategy will need to 
be developed in order ensure that the residual biases in these surveys fall 
below the statistical errors.
\end{abstract}
\maketitle

\section{Introduction}

%------------------------------------------------------

Weak gravitational lensing is a potentially powerful probe of cosmology 
(e.g., Refs.~\cite{Albrecht:2006um,hoekstra_jain08,weinberg_etal12,hu99,hu_tegmark99,refregier_etal04,hu_jain04})
\footnote{This application of lensing goes back more than forty years (e.g., Ref.~\cite{kristian67})}. 
Imaging surveys such as the Dark Energy Survey (DES) and, in the longer term, 
the surveys of the Large Synoptic Survey Telescope (LSST), the European 
Space Agency's Euclid satellite, and the Wide Field Infra-Red Survey Telescope (WFIRST) 
expect to measure the power spectrum of cosmological weak lensing with sufficient precision 
to improve constraints on dark energy dramatically.  However, a number of 
sources of systematic error must be controlled in order to achieve 
these goals. From a theoretical perspective, it is necessary to predict matter 
power spectra with precisions of better than one percent over a wide range of 
scales \cite{huterer_takada05,hearin_etal12}. This is a challenging goal, but 
significant progress has been realized utilizing N-body simulations containing 
only dark matter \cite{heitmann_etal09,lawrence_etal10,heitmann_etal10,Eifler:2011}. The 
largest remaining challenge to this goal is to account for the influence of 
the baryonic component of the universe in these predictions.  Baryonic effects 
have been shown to alter lensing power spectra significantly on small scales 
\cite{zhan_knox04,white04,jing_etal06,Rudd:2007zx,guillet_etal10,semboloni_etal11}.  
This theoretical systematic error associated with baryonic processes is sufficient 
to cause large systematic errors in inferred dark energy parameters if unaccounted 
for \cite{Zentner:2007bn,hearin_zentner09,semboloni_etal11,hearin_etal12}, though 
Ref.~\cite{huterer_white05} explored methods to cull data in order to 
protect against scale-dependent uncertainties in predicted power spectra. 
In the present work, we assess a proposal to mitigate dark energy biases induced 
by baryonic effects using a method proposed in Ref.~\cite{Zentner:2007bn}.

\citet{Rudd:2007zx} recognized changes in the internal structures of dark matter 
halos as the cause of the largest alterations to lensing spectra in baryonic 
simulations (a result confirmed in Refs.~\cite{guillet_etal10,semboloni_etal11}).  
Consequently, \citet{Zentner:2007bn} suggested a strategy to 
mitigate baryonic effects in forthcoming lensing analyses. 
\citet{Zentner:2007bn} proposed altering the canonical relationship 
between halo mass and halo concentration (e.g., \cite{munoz-cuartas_etal11,prada_etal12}) 
to account for the matter redistribution precipitated by baryonic effects, 
as this enables the simulations of Ref.~\cite{Rudd:2007zx} to be modeled successfully.  
\citet{Zentner:2007bn} then suggested introducing additional parameter freedom into 
the concentration-mass relation, a necessity because this relation cannot be unambiguously predicted 
due to the uncertainties in baryonic processes, and fitting the data simultaneously for the 
parameters that quantify the mass-concentration 
relation and the cosmological parameters. 
The value of this strategy is that it can reduce systematic errors (or {\em biases}) 
in dark energy parameters to acceptable levels, while increasing the 
statistical errors on dark energy parameters by only $\sim 10-40\%$, depending upon the experiment and the 
complexity of the mass-concentration relation \cite{Zentner:2007bn} 
(a similar argument can be made for modified gravity \cite{hearin_zentner09}).

\Sfig{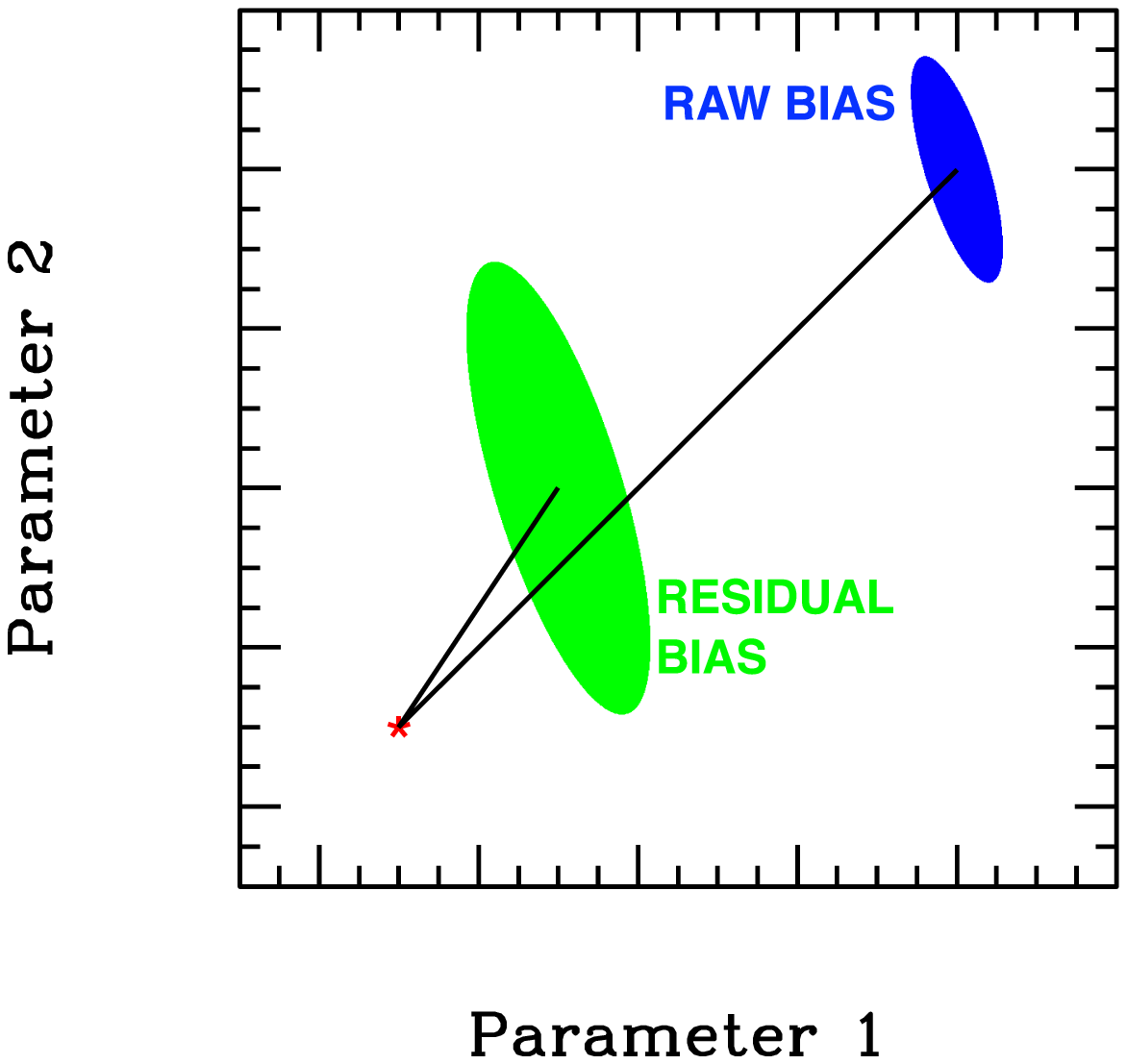}{Cartoon view of the effect of bias on cosmological 
parameters (here called ``Parameter 1'' and ``Parameter 2'').
The true value of the parameters is given by the red star.
If the bias (in our case the effect of baryons on the weak lensing power spectrum) 
is not accounted for, the allowed region in parameter space
will be given by the shaded blue region at the top right. The parameters will be offset 
from their true values, or {\em biased}. We call the offset in this case 
the ``raw bias.'' If one attempts to mitigate the bias by introducing new 
parameters (in our case allowing for a varying mass-concentration relation), 
the allowed region will shift to that given by the shaded green contour at the lower left. 
The errors are now larger due to the increased number of parameters used 
in the fit, but the offset, which we refer to as the 
``residual bias,'' is much smaller than the raw bias.}

Our aim here is to test this mitigation strategy more extensively. 
We wish to determine if this algorithm will extract 
cosmological parameters successfully from upcoming survey data.  
{\em Successfully} here has a specific and technical meaning, 
a cartoon version of which is illustrated in Fig.~\ref{fig:cartoon}.
First, success demands that biases in the cosmological parameters due 
to inaccuracies in theoretical models should be small. There are two biases at play here,
the ``raw bias'' (the offset of the smaller contour in Fig.~\ref{fig:cartoon}) 
before any mitigation is applied and the ``residual bias'' (offset of the 
larger contour in Fig.~\ref{fig:cartoon}) which remains after
fitting for the new free parameters. Ideally the residual bias 
will be much smaller than the raw bias;
for the method to be truly effective, the residual bias 
should be smaller than the statistical error.
Second, success requires that 
the additional parameter freedom introduced 
by the model should not inflate the error bars 
on cosmological parameters so much as to markedly reduce the 
constraining power of the experiment. At minimum, the increase 
in the statistical error bars due to additional parameters should not be so 
large as to nullify the reduction in the systematic errors.

To carry out this test, we use the results from the OverWhelmingly Large Simulations 
(OWLS)~\cite{Schaye:2009bt,vanDaalen:2011xb,semboloni_etal11} as mock data.  
The OWLS suite consists, 
in part, of a set of ten simulations, each with the same initial conditions evolved 
in the context of the same cosmology. One simulation treats only dark matter, while 
the other nine model baryonic processes using different effective models.  We proceed 
by assuming that each one of the OWLS simulations, in turn, produces the {\em true} 
matter power spectrum. We fit each of the OWLS predictions for lensing power 
spectra with our mitigation model, including nuisance parameters. We compute 
residual differences in power spectra between our fits and the OWLS predictions 
and use these differences to estimate the biases in dark energy equation of 
state parameters that would be realized after applying the mitigation scheme.
We repeat this analysis in the context of two distinct imaging surveys. 
The first survey we consider has the precision expected from DES. 
The second survey we consider represents more long-term, 
Stage IV\footnote{Using the classification scheme of the 
Dark Energy Task Force \cite{Albrecht:2006}, within which DES 
would be a Stage III experiment.}
surveys, such as may be conducted by LSST or Euclid.

We will show that baryonic effects may reasonably 
lead to raw biases as large as $\sim 2\sigma-6\sigma$ 
(where $\sigma$ represents the marginalized 
statistical error) on dark energy equation of state parameters if 
unaccounted for in the analysis of DES-like data. The size of the bias 
depends upon the range of multipoles used in the analysis and the 
baryon model. This broadly confirms 
prior estimates \cite{Zentner:2007bn,hearin_zentner09,semboloni_etal11}. 
We will then show that this mitigation scheme can render systematic 
errors sufficiently small, so as to suggest concentration fitting as an 
attractive strategy for the cosmological analysis of lensing power 
spectra from DES. In all cases that we consider, the residual biases 
remain $\lesssim 0.5\sigma$ and can be kept $\lesssim 0.1\sigma$ if 
the range of scales included in the cosmological analysis is restricted 
to $\ell \lesssim 2000$, though restricting scales comes at a non-negligible 
cost in statistical error.

For Stage IV experiments with wide sky coverage, such as LSST or Euclid, 
the conclusion is slightly more complicated. 
Absent any mitigation scheme for baryonic process 
such future experiments may be subject to raw biases ranging 
from $\sim 1.5\sigma$ to as large as ten times the statistical error or more. 
The broad range reflects the differences 
from one OWLS simulation to the next. However the largest of these 
biases are unlikely to be the product of any actual analysis. 
It seems more likely that the team undertaking the analysis 
will notice that all models provide poor fits to the data 
using some ``goodness-of-fit'' criterion. Nevertheless, it remains 
imperative to understand the reasons for the poor fits. Our analysis 
suggests that concentration fitting may reduce systematic errors 
on dark energy equation of state parameters due to baryonic effects 
to $\lesssim 1.6\sigma$ in the worst case and $\lesssim 0.5\sigma$ in 
six of the nine simulations we have analyzed. The concomitant 
increase in the statistical errors is $\lesssim 30\%$. 
While concentration fitting does alleviate 
biases in this case, a more sophisticated analysis may be 
necessary for data of the quality expected from Stage IV experiments.

The remainder of this manuscript is organized as follows. In the 
following section, we describe the lensing power spectra from which 
we aim to infer cosmological parameters, 
the details of our modeling procedure, 
and the cosmological parameters that we consider.  
We also discuss the Fisher matrix method for estimating 
statistical and systematic errors 
in model parameters. In \S~\ref{section:sims}, we 
describe the OWLS simulations and show the differences 
in lensing power spectra predicted by several of the 
simulations in the OWLS suite.  We describe our 
simple mitigation model in \S~\ref{section:fit}.  
Our results for the statistical and systematic 
errors on dark energy parameters are given in 
\S~\ref{section:results}, where we address a DES-like 
experiment, and a future LSST- or Euclid-like experiment 
in turn. We summarize our results and present our 
conclusions in \S~\ref{section:conclusions}.

%-------------------------------------------------------------------
\section{Preliminaries}       % Enter section title between curly braces
\label{section:preliminaries}
%-------------------------------------------------------------------

\subsection{Weak Lensing Observables}

We consider cosmological parameter inference using measurements of cosmic shear 
from large-scale imaging surveys.  We assume that each galaxy has a well-characterized 
photometric redshift estimate, so that the source galaxies can be binned in $\ntomo$ photometric 
redshift bins.  We infer cosmological parameters from the 
$\nobs = \ntomo (\ntomo + 1)/2$ number density-weighted angular convergence 
power spectra and cross spectra among the galaxies in each of the redshift bins, 
\be
\label{eq:pkij}
\Pktom{i}{j}(\ell) = \int \dd z  \frac{\wlw{i}(z)\wlw{j}(z)}{H(z)\Da^2(z)}P_{\delta}(k=\ell/\Da,z).
\ee
In Eq.~\ref{eq:pkij}, $H(z)$ is the Hubble expansion rate, $\Da(z)$ is the 
angular diameter distance to redshift $z$, $P_{\delta}(k,z)$ is the three-dimensional 
matter power spectrum at wavenumber $k$ and redshift $z$, $\wlw{i}(z)$ are the $\ntomo$ 
lensing weight functions, and the lower-case Latin indices indicate the redshift 
bins (e.g., index $\mathrm{i}$ runs from $1$ to $\ntomo$).  
The lensing weight functions are 
\be
\label{eq:wi}
\wlw{i}(z) = \frac{3}{2}\Omegam H_0^2 (1+z) \Da(z) 
\int \dd z' \frac{\Da(z,z')}{\Da(z')}\ \frac{\dd n_{\mathrm{i}}}{\dd z'}, 
\ee
where $\dd n_{\mathrm{i}}/\dd z$ is the redshift distribution of source galaxies 
in the $\mathrm{i}^{\mathrm{th}}$ redshift bin, $H_0$ is the present Hubble rate, 
and $\Da(z,z')$ designates the angular diameter distance between redshifts 
$z$ and $z'$.

The observed spectra $\Pkobs{i}{j}$, consist of terms due to signal ($\Pktom{i}{j}$) 
and noise, 
\be
\label{eq:pobs}
\Pkobs{i}{j}(\ell) = \Pktom{i}{j}(\ell) + n_{\mathrm{i}} \delta_{\mathrm{ij}} \gi ,
\ee
where $n_{\mathrm{i}}$ is the surface density of source galaxies in redshift bin i, 
$\gi$ is the intrinsic source galaxy shape noise for each shear component, and 
$\delta_{\mathrm{ij}}$ is the Kronecker delta symbol. The covariance among 
observables is 
\be
\label{eq:covar}
\cov[ \Pktom{i}{j}(\ell), \Pktom{k}{l}(\ell )] = \Pkobs{i}{k}\Pkobs{j}{l} + \Pkobs{i}{l}\Pkobs{j}{k},
\ee
assuming Gaussian statistics. Over the range of scales we consider, 
the Gaussian approximation is reasonable (e.g., Ref.~\cite{semboloni_etal06}) 
and greatly simplifies the analysis. Moreover, it is a 
conservative assumption for our purposes because 
adopting non-Gaussian covariance generally renders statistical errors 
larger and diminishes the relative importance of the systematic errors we consider. 
Throughout this study, we adhere to a common convention by taking $\sqrt{\gi} = 0.2$.

\subsection{Survey Characteristics and Cosmological Parameters}

We consider cosmological constraints from two representative surveys.  
The first experiment we consider is based on the specifications of the 
Dark Energy Survey (DES)\footnote{{\tt http://darkenergysurvey.org}}. 
DES is an example of a near-term, ``Stage III'' project that will exploit 
cosmic shear measurements to derive constraints on dark energy parameters. 
We model DES by taking a fractional sky coverage of $\fsky=0.12$, 
corresponding to approximately $5000\ \mathrm{deg.}^2$, and a total 
surface density of imaged galaxies of $N^{\mathrm{A}}=15\ \mathrm{arcmin}^{-2}$. 
This choice is optimistic, but it is a conservative assumption for our purposes 
as smaller statistical error bars set a more stringent requirement for the mitigation 
of systematic errors. We take the DES redshift distribution of source galaxies 
from the DES Blind Cosmology Challenge (BCC) simulation. The DES BCC 
simulation comprises 5000 square degrees of simulated shear maps and is 
tuned to match the expected observational characteristics of the DES mission.  
We divide the source galaxies into $\ntomo=5$ 
redshift bins such that 20\% of the total number of observed 
galaxies are placed in each bin, this gives $\nobs=15$ distinct convergence 
spectra.  Binning more finely in redshift does not alter our results, 
in accord with prior studies \cite{ma_etal06,hearin_etal12}. 
Throughout the remainder of this paper, 
we will refer to results based on these survey specifications 
by the name ``DES.''

In addition to DES, we will estimate the potential influences 
of baryonic physics on dark energy 
constraints from long-term future experiments,
categorized as ``Stage IV'' experiments in the report of 
the Dark Energy Task Force \cite{Albrecht:2006}. Examples of 
potential Stage IV experiments that will explore cosmological 
constraints from weak gravitational lensing are the 
Large Synoptic Survey Telescope (LSST, Ref.~\cite{lsst_book})\footnote{{\tt http://www.lsst.org}} 
or the European Space Agency's Euclid\footnote{{\tt http://sci.esa.int/euclid}} 
project \cite{laureijs_etal12}.  
We characterize these experiments by a fractional sky coverage of 
$\fsky=0.5$ and a number density of source galaxies of 
$N^{\mathrm{A}} = 30\ \mathrm{arcmin}^{-2}$
\footnote{The sky coverage of Euclid will likely be closer 
to $\fsky \approx 1/3$ \cite{laureijs_etal12}.}.  Again, these choices 
are optimistic, but they {\em maximize} the relative importance of 
the systematics we aim to militate against, so they are conservative 
choices for our purposes. We model the redshift 
distribution of source galaxies in these long-term surveys 
as $\dd n/\dd z \propto z^2 \exp( -(z/z_0)^{1.2})$, 
with $z_0 \simeq 0.34$ to give a median redshift $z_{\mathrm{median}}=1$.  
This choice is based on the approximate, 
observed distribution of high-redshift galaxies \cite{deep2}.  
As with DES, we place the source galaxies into $\ntomo=5$ 
redshift bins so that the 20\% of the galaxies fall into 
each bin.  We refer to results with these specifications 
as ``Stage IV'' results.

We consider cosmologies defined by seven parameters, three of 
which describe the dark energy.  The parameters that describe 
the dark energy are the contemporary dark energy density in units 
of the critical density, $\Omega_{\mathrm{DE}}$, and the two parameters 
$\wzero$ and $\wa$ that specify a dark energy equation of state 
that varies linearly with scale factor, $w(a)=\wzero+\wa(1-a)$.  
The parameters of our fiducial cosmology are fixed to match the 
cosmological parameters assumed in the OWLS simulation program \cite{Schaye:2009bt}.  
The parameters specified in the OWLS program are 
the matter density, $\omegam = 0.1268$, 
the baryon density, $\omegab = 0.0223$, 
the scalar spectral index, $n_{\mathrm{s}}=0.951$, 
the amplitude of curvature fluctuations on 
a scale of $k=0.05\, \mathrm{Mpc}^{-1}$, 
$\Delta^2_{\mathcal{R}}=1.9\times 10^{-9}$ 
(we actually vary $\ln \Delta^2_{\mathcal{R}}$ about this value), 
$\Omega_{\mathrm{DE}}=0.762$, $\wzero=-1$, and $\wa=0$.  
These parameter values imply that the root-mean-square 
matter density fluctuation on a scale of $8\ h^{-1}$~Mpc  
is $\sigma_8 = 0.74$.  We include prior constraints on 
these parameters that reflect expected limits from 
the Planck cosmic microwave background anisotropy measurements in 
all of our calculations. The Planck prior matrix that we use was computed in Ref.~\cite{hu_etal06}. 
In addition to these seven cosmological parameters, 
we introduce three other parameters, described in \S~\ref{section:fit}, 
that account for baryonic effects.

\subsection{Methodology}

In principle, we propose to assess the effectiveness of the mitigation approach proposed in Ref.~\cite{Zentner:2007bn} 
using the following steps. 
\begin{enumerate}
\item Take the lensing spectra predicted by one of the OWLS simulations as mock data.
\item Fit the mock data to a model by varying 7 cosmological parameters and 
determine the statistical errors on the cosmological parameters. This model 
{\em does not} include the effects of baryons.
\item Determine the {\em raw bias} as the difference between the resulting best-fit 
dark energy parameters and the ``true'' input parameters used to generate the OWLS simulations. 
\item Fit the mock data again to a model with those same 7 cosmological parameters as well as 
3 additional parameters that account for baryonic effects.
\item Determine the {\em residual bias} as the difference between this second fit 
and the ``true'' values of the dark energy parameters used in the OWLS simulations.
\item Compare the size of the error bars in both cases to see the amount by which 
the errors are inflated as a result of the new degrees of freedom. 
\item Repeat for each of the OWLS simulations to arrive at nine distinct assessments.
\end{enumerate}
In practice, going through this entire process for all the cases of interest would be extremely 
time consuming, because fitting for the cosmological and concentration parameters 
in the multi-dimensional parameter space that we explore is a 
computationally-expensive task. Instead, we proceed using an approximation, 
based on both direct fitting for model parameters and Fisher matrix (described below) 
estimates for the statistical and systematic errors in model parameters.
However, it is important to stress that we are assessing the residual bias that will 
ensue if analysts follow the mitigation strategy proposed in Ref.~\cite{Zentner:2007bn} 
on upcoming data sets.

In order to limit computational effort, we use the Fisher information matrix 
to assess the constraining power of these $\nobs$ observable spectra.  We 
assume that the spectra are independent, Gaussian random variables at 
each multipole, so that the Fisher matrix may be written as 
\begin{eqnarray}
\label{eq:Fisher}
\fish_{\mathrm{AB}} & = & \fish^{\mathrm{P}}_{\mathrm{AB}} \nonumber \\ 
              & + & \sum_{\ell=\ell_{\mathrm{min}}}^{\ellmax} (2\ell + 1)\fsky 
\sum_{\mathrm{i}=1}^{\ntomo}\sum_{\mathrm{j}=\mathrm{i}}^{\ntomo} 
\sum_{\mathrm{k}=1}^{\ntomo} \sum_{\mathrm{l}=\mathrm{k}}^{\ntomo} \nonumber \\
              & \times & \frac{\partial \Pktom{i}{j}(\ell)}{\partial p_{\mathrm{A}}}\, 
\cov^{-1}[ \Pktom{i}{j}(\ell), \Pktom{k}{l}(\ell) ]\, \frac{\partial \Pktom{k}{l}}{\partial p_{\mathrm{B}}}.
\end{eqnarray}
The matrix $\fish^{\mathrm{P}}_{\mathrm{AB}}$ represents the prior constraints on the 
cosmological parameters, $p_{\mathrm{A}}$ are the parameters of the model, 
and $\cov^{-1}[ \Pktom{i}{j}, \Pktom{k}{l} ]$ is the inverse of the covariance 
matrix between observables.  The upper-case Latin indices signify model parameters.  
The parameter $\fsky$ specifies the fraction of 
sky observed in the survey, and the sum runs over multipoles from $\ell_{\mathrm{min}}$ 
to $\ellmax$.  We take $\ell_{\mathrm{min}}=2\fsky^{-1/2}$; however all of the constraints 
we consider are dominated by multipoles significantly higher than $\ell_{\mathrm{min}}$ 
so that this choice is inconsequential.  We take $\ellmax = 5000$ throughout most of 
our study so as to remain in a regime in which a number of 
simplifying assumptions are approximately valid 
(e.g., Refs.~\cite{Hu:2000ax,white_hu00,cooray_hu01,vale_white03,dodelson_etal06,semboloni_etal06}), 
but we explore other choices of maximum multipole. 
Including such high multipoles in our analysis may well be overly optimistic. 
However, using higher multipoles (smaller scales) in the cosmological analysis 
results in greater constraining power, so it is interesting to determine 
the utility of our mitigation scheme out to relatively high multipoles. 
The Fisher matrix approximates the covariance among model parameters at the 
maximum of the likelihood, so that the error in the estimate of the 
$\mathrm{A}^{\mathrm{th}}$ parameter can be approximated as 
$\sigma(p_{\mathrm{A}})=\sqrt{\fish^{-1}_{\mathrm{AA}}}$, after marginalizing 
over the remaining parameters.

The Fisher matrix formalism provides 
a straightforward estimate of parameter biases due to undiagnosed, systematic 
offsets in observables.  Let $\Delta \Pktom{i}{j}$ represent the difference 
between the true observable and the observable perturbed due to some systematic 
effect.  A Taylor expansion about the maximum likelihood gives an estimate of 
the systematic error contribution to model parameter $p_{\mathrm{A}}$ due to 
the systematic offsets in observables \cite{knox_etal98}: 
\begin{eqnarray}
\label{eq:fisherbias}
b(p_{\mathrm{A}}) & = & \sum_{\mathrm{B}} \fish^{-1}_{\mathrm{AB}} \sum_{\ell} (2\ell + 1)\fsky \nonumber \\ 
                & \times & \sum_{\mathrm{i,j,k,l}} 
\Delta \Pktom{i}{j}\, \cov^{-1}[ \Pktom{i}{j}(\ell), \Pktom{k}{l}(\ell) ]\, 
\frac{\partial \Pktom{k}{l}}{\partial p_{\mathrm{B}}}.
\end{eqnarray}
The sums over observable (lower-case Latin) indices in Eq.~(\ref{eq:fisherbias}) 
have the same form as those in Eq.~(\ref{eq:Fisher}), though we have written them as a 
single sum for brevity.

The practical strategy that we implement in an effort to limit 
computational expense is a modification to the ideal strategy that 
we would, in principle, pursue as describe above. 
We {\em repeal and replace} {\tt Step 2} through {\tt Step 5} 
with the following steps.
\begin{enumerate}
\item[2$'$.] Determine the {\em raw bias} using the Fisher matrix 
relation [Eq.~(\ref{eq:fisherbias})] with the systematic offsets 
in power spectra given by the difference between the OWLS baryonic 
simulation and the fiducial OWLS simulation that treats dark matter only.
\item[3$'$.] Fit the 3 parameters of the concentration-mass relation of halos 
to the OWLS convergence power spectra within a {\em fixed} cosmological model. 
This results in a best-fit concentration-mass relation that best 
describes the OWLS simulations within the true underlying OWLS cosmological model.
\item[4$'$.] Compute the residual differences between the predicted convergence spectra from the OWLS 
simulations and those predicted by the best-fit concentration-mass model with cosmological parameters 
fixed to the OWLS cosmological model. These residual differences, $\Delta \Pktom{i}{j}$, will 
give rise to systematic errors in inferred cosmological parameters.
\item[5$'$.] Use the Fisher matrix formalism [particularly Eq.~(\ref{eq:fisherbias})] to 
estimate the {\em residual bias} from $\Delta \Pktom{i}{j}$ 
after allowing for parametric freedom in the concentration-mass relation. 
This level of bias will not be removed by fitting for halo structure 
and will remain in the error budget on cosmological parameters.
\end{enumerate}

This strategy enables us to estimate the efficacy of the 
baryonic physics mitigation proposal. Additionally, it requires 
explicitly fitting for only the concentration-mass relation 
(a 3-dimensional subspace of the full parameter space), so that 
the computational effort required is significantly less than 
would be required to fit for halo structure and cosmology 
simultaneously. The merit of this approach is that the fit 
allows us to assess the fidelity with which we can model lensing 
power spectra including baryonic effects and it brings our fiducial 
model closer to the true, underlying model prior to applying the 
Fisher formalism in step $5'$. 
This modified procedure introduces a possible source of confusion.  One 
might think that our final statistical and systematic error estimates 
do not include the covariance between cosmological and concentration 
parameters. We emphasize that this is not the case.  The Fisher matrix 
formalism for computing biases in model parameters accounts for the 
covariance among all the model parameters.  The limitation 
is the standard caveat that the Fisher matrix is a first-order 
approximation about the maximum likelihood.

%-------------------------------------------------------------------
\section{Spectra from Simulations}
\label{section:sims}
%----------------------

We estimate the influence of baryonic processes on weak lensing power spectra using the 
OverWhelmingly Large Simulations (OWLS)~\cite{Schaye:2009bt,vanDaalen:2011xb,semboloni_etal11}.  
The OWLS suite includes a set of ten simulations set within the same cosmological model 
that evolve from the same initial conditions to redshift $z=0$.  One of these simulations, the ``DMONLY'' 
simulation, treats the gravitational evolution of dissipationless dark matter only.  This 
type of dissipationless dark matter simulation is similar to the vast majority of simulations 
that have been used to model the observations of contemporary and future surveys.  The 
remaining nine simulations include the baryonic component of the Universe along with 
various effective models for baryonic gas cooling, star formation, 
and a number of feedback processes.  It is not feasible to 
simulate these processes directly, so baryonic simulations rely on a variety of effective 
models for these processes.  Effective models for baryonic processes remain quite 
uncertain and it is not possible to produce a definitive prediction for the influences 
of baryonic processes on observables, such as convergence power spectra.  The utility 
of a simulation suite such as OWLS, is that it provides a range of distinct, but plausible 
predictions for observables so that systematic errors induced by our ignorance of 
baryonic physics can be estimated. An important advantage of the test that we present 
here is that we are applying a mitigation strategy developed on 
the simulations of \citet{Rudd:2007zx} to an independent 
set of simulations that were performed 
using markedly different simulation strategies. The details of the OWLS simulations have been 
given in Refs.~\cite{Schaye:2009bt,owls_agn,owls_se,owls_rc,owls_sn,owls_sf}, while the OWLS power spectra were the 
subject of Ref.~\cite{vanDaalen:2011xb}, to which we 
refer the reader as these details are not of immediate importance in the present paper.
Follow up studies by \citet{mccarthy_etal10} and \citet{mccarthy_etal11} suggest that 
the properties of galaxies and hot gas in galaxy groups are modeled most reliably 
in the ``AGN'' simulation, which includes strong feedback from active galactic nuclei.

We regard each of the nine baryonic simulations separately as a potential realization 
of the effects of baryons on convergence power spectra.  Accordingly, we treat the 
convergence spectra from {\em each} of the OWLS baryonic simulations as ``observed'' spectra 
that must be modeled faithfully in order to extract reliable constraints.  
The aim is to test whether a specific strategy to mitigate the influence of baryonic 
processes on dark energy constraints can be applied to a variety of distinct predictions 
successfully.  {\em Success} in this context means that the mitigation procedure 
renders the biases in dark energy parameters significantly smaller than the 
expected statistical errors.  If a {\em single} mitigation strategy were to achieve 
the requisite reduction in dark energy parameter biases for {\em all} plausible 
simulations, it would be a strong indication that the mitigation strategy may be 
applied to observational data to limit systematic errors on dark energy parameters 
associated with the influences of baryons.

\citet{vanDaalen:2011xb} used the OWLS simulations to study the effects of baryonic physics on the matter 
power spectrum.  We use the 3D matter power spectra, $P_{\delta}(k,z)$, provided in Ref.~\cite{vanDaalen:2011xb} 
for the OWLS simulations to estimate convergence power spectra using Eq.~(\ref{eq:pkij}).  In practice, 
the tabulated matter power spectra from Ref.~\cite{vanDaalen:2011xb} cannot be used directly to 
predict convergence power spectra.  Due to computational limitations, the OWLS simulation volumes 
are relatively small (cubic boxes $L=100\ h^{-1}\mathrm{Mpc}$ on a side), and are subject to 
significant sample variance and finite volume effects on large scales.  In order to overcome 
these drawbacks, we utilize the OWLS $P_{\delta}(k,z)$ tables directly for wavenumbers 
$k > 0.314\ h\mathrm{Mpc}^{-1}$.  The OWLS spectra are reliable for 
$k < 10\ h\mathrm{Mpc}^{-1}$ \cite{vanDaalen:2011xb},  
which is sufficient for our purposes \cite{hearin_etal12}.  For wavenumbers $k < 0.314$, we use the 
halo model as implemented in Ref.~\cite{Zentner:2007bn} to estimate the matter power spectrum.  
We multiply the halo model power spectra by a correction factor that ensures that the two 
spectra agree at $k=0.314\ h\mathrm{Mpc}^{-1}$.  In the OWLS simulations, 
baryonic effects induce changes in power spectra of 
$\lesssim 1\%$ on scales $k \lesssim 0.314\ h\mathrm{Mpc}^{-1}$. 
An important caveat to our approach is that assume that the effects of 
baryons on scales $k \lesssim 0.3\ h\mathrm{Mpc}^{-1}$ are insignificant.

%-----------------------------------
% simulation residuals
\begin{figure}[t!]
\includegraphics[width=9cm]{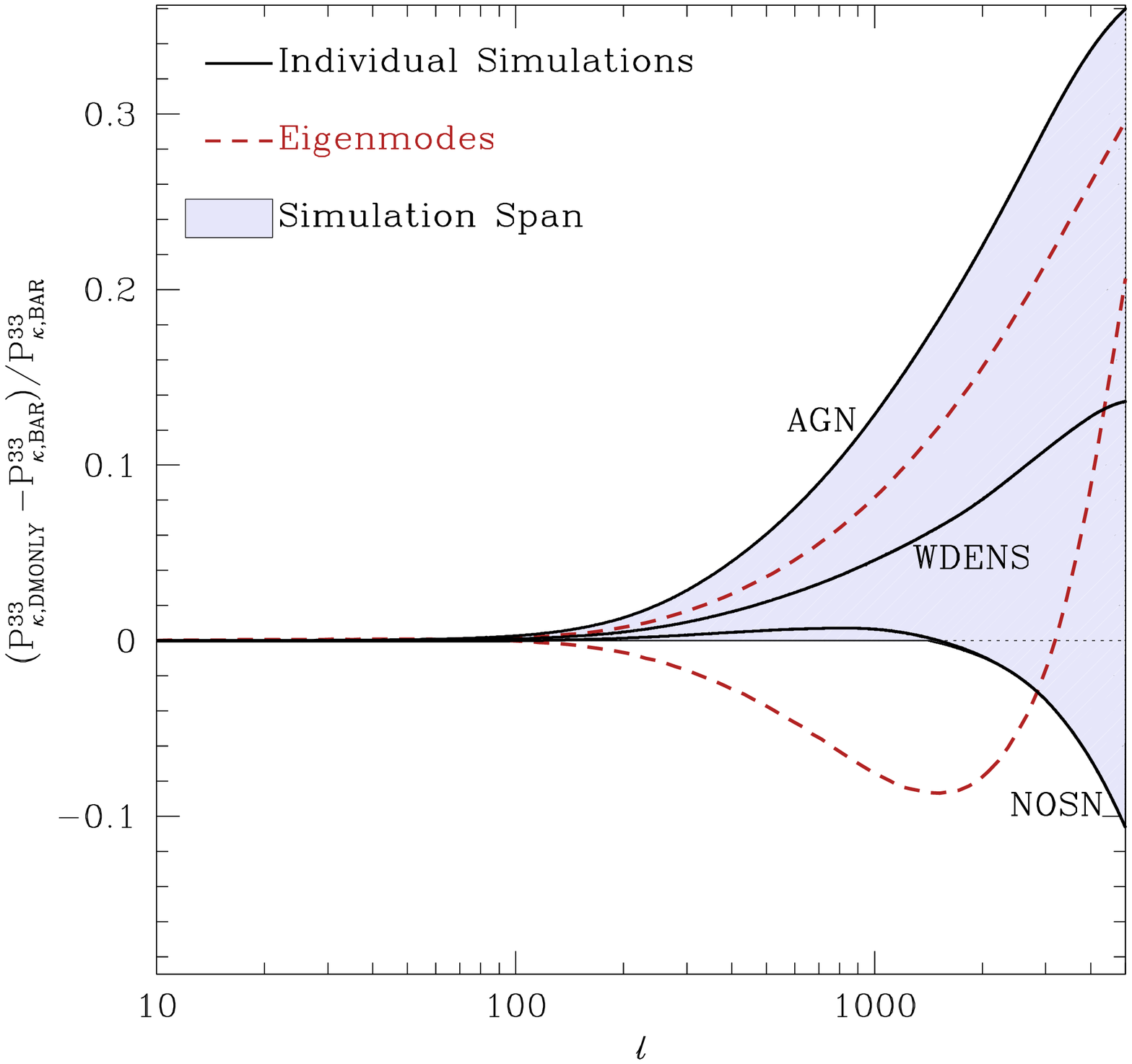}
\caption{ 
Convergence power spectrum residuals between the DMONLY simulation and baryonic simulations 
(denoted ``BAR'' in the vertical axis label).  For simplicity, this figure shows only the 
residual for the auto power spectra in the third tomographic bin ($0.5025 \le z \le 0.6725$), 
$\Pktom{3}{3}$, assuming a DES analysis. 
Fractional residuals of the other fourteen observables display similar features. 
The shaded band covers the region spanned by the residuals of {\em all} of the OWLS 
baryonic simulations.  The three {\em thick, solid} lines show three specific simulations 
that contribute to this band, namely, the AGN (bounding the shaded region above), 
NOSN (bounding the shaded region below for $\ell \gtrsim 2000$, and WDENS (intermediate) 
simulations. The residuals at different multipoles are highly correlated.  
The {\em dashed} lines show the principal modes of the residuals that have 
the highest (upper) and second-highest (lower) variance.  
These modes account for over $90\%$ of the variance among the spectra 
and demonstrate the correlated manner in which baryons alter lensing 
power spectra.
}
\label{fig:resband}
\end{figure}
%------------------------------------

Figure~\ref{fig:resband} shows the fractional differences between the 
convergence power spectra predicted by the baryonic simulations 
compared to the DMONLY simulation.  For simplicity, we show only 
the power spectrum in our third DES redshift bin, $\Pktom{3}{3}$ 
($0.5025 \le z < 0.6725$ for our DES model). 
The predictions for the other fourteen 
observables show similar features.  Fig.~\ref{fig:resband} shows 
these residuals in several distinct ways.  The shaded band is the 
envelope of the power spectrum residual constructed from all nine 
of the baryonic simulations.  The residual power spectra from three  
specific simulations, namely AGN (bounding the shaded region above), 
NOSN (bounding the shaded region below for $\ell \gtrsim 2000$), 
and WDENS (intermediate), are shown as solid lines.

%-----------------------------------
% fit residuals
\begin{figure}[t!]
\includegraphics[width=9cm]{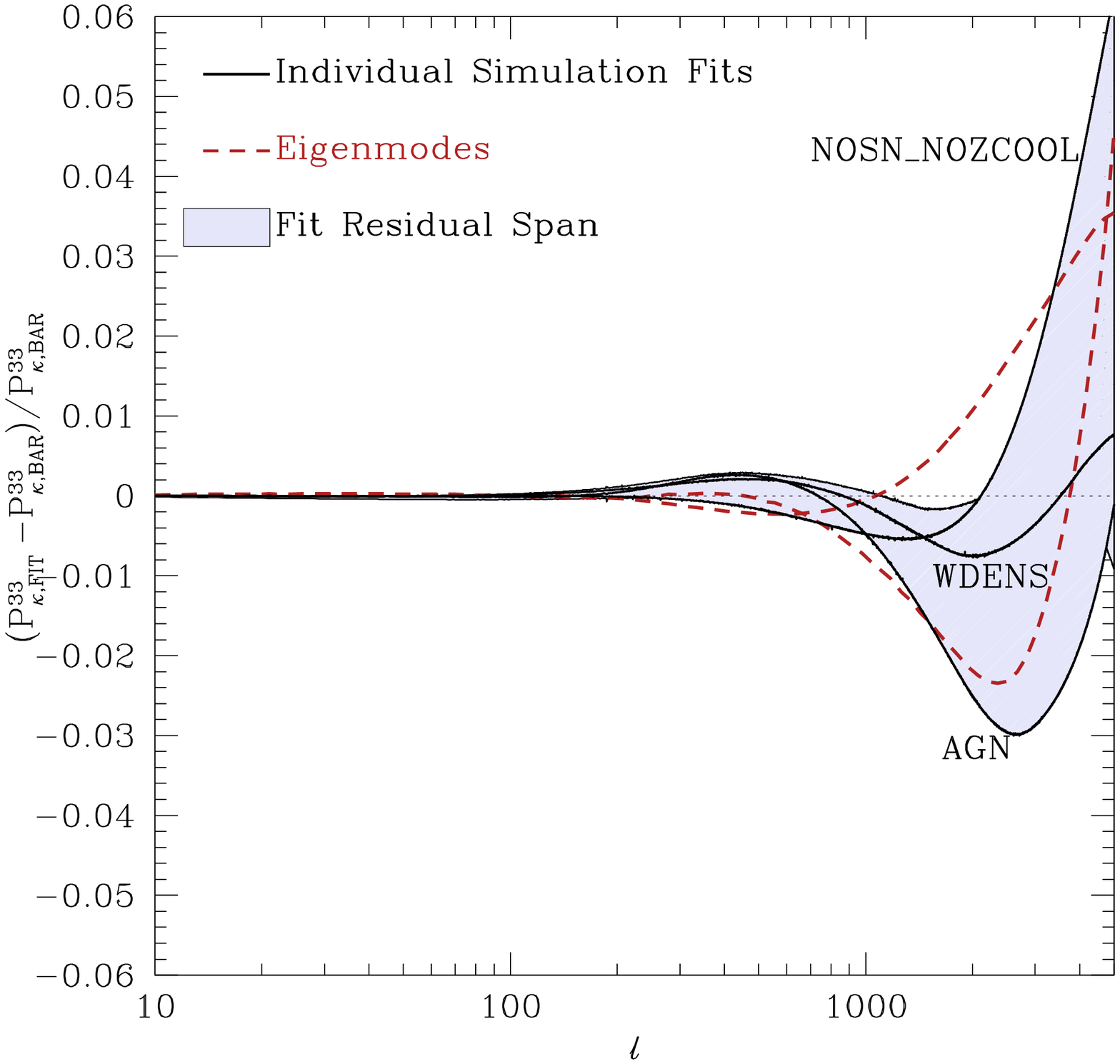}
\caption{ 
Convergence power spectrum residuals between our halo model fits to the spectra 
predicted by the baryonic simulations and the baryonic simulation predictions themselves.  
The shaded band shows the envelope containing the fit residuals for all 
of the OWLS baryonic simulations, analogous to the shaded band in 
Fig.~\ref{fig:resband}. As in Fig.~\ref{fig:resband}, the {\em thick, solid} lines 
show examples of the fit residuals from specific cases, namely the 
AGN (lowest at high multipole), NOSN\_NOZCOOL (highest at high multipole, lowest 
at $\ell \lesssim 1000$), and WDENS (intermediate) simulations. It is evident 
that the envelope containing all residuals is comprised by different simulations 
over different multipole ranges. The {\em dashed } lines show the 
two principle modes of the residuals with the highest variance.  
These represent correlated failure modes of the fitting procedure. 
This figure is to be compared to Fig.~\ref{fig:resband}, but 
notice that the ranges on the ordinal axes are considerably different 
in the two figures.  The halo model fits typically reduce power 
spectrum residuals by more than an order of magnitude.
}
\label{fig:fitband}
\end{figure}
%-----------------------------------

As is evident in the AGN, NOSN, and WDENS models, 
the deviations in the spectra predicted by the baryonic simulations 
differ from the DMONLY simulation in a way that is correlated 
from multipole to multipole.  Accordingly, the shaded bands in Fig.~\ref{fig:resband} 
represent the {\em envelope} of deviations, while an individual 
spectrum will not range over the shaded area.  To represent 
the typical shapes of the baryonic simulation spectra, we have 
performed the following exercise.  We have computed the covariance 
among the distinct spectra of the baryonic simulations, 
$\cov[\Pktom{3}{3}(\ell),\Pktom{3}{3}(\ell')] = N^{-1}\sum_{i=1}^{N} 
\left( \Pktom{3}{3,i}(\ell) - \left\langle \Pktom{3}{3}(\ell) \right\rangle \right)
\left( \Pktom{3}{3,i}(\ell') - \left\langle \Pktom{3}{3}(\ell') \right\rangle \right)$, 
where $i$ is an index designating the simulation, $N=9$ is the number 
of baryonic simulations, and $\langle \Pktom{3}{3}(\ell) \rangle$ is the 
average of the spectra from all of the simulations. We then 
diagonalized the covariance matrix, $\cov[\Pktom{3}{3}(\ell),\Pktom{3}{3}(\ell')]$.  
The eigenvectors of the covariance matrix represent the principal modes of variation 
of the power spectra and the eigenvalues represent the variance 
accounted for by the corresponding eigenvectors.  The dashed 
lines in Fig.~\ref{fig:resband} are the two eigenvectors 
corresponding to the first- and second-largest eigenvalues.  
These principal modes account for over $90\%$ of 
the variance among the spectra and illustrate the correlated 
manner in which the baryonic simulation spectra may differ 
from the DMONLY predictions.

%--------------------------------------------------------------
\section{Fitting for Baryonic Effects with Halo Concentrations }
\label{section:fit}
%----------------

%%%%%%%%%%%%%%%%%%%%%%%%%%%%%%%%%%%%%%%%%%%%%%%%%%%%%%%%%%%%%%%%%%%%%%%%%%%%
\begin{table}[t]
\caption{
Concentration parameters that best-fit the OWLS simulation 
spectra. The best-fit values are given for the normalization 
of the concentration relation $c_0$, in units of the normalization 
for the DMONLY simulation, the power-law index specifying the 
mass dependence $\alpha$, and the power-law index of the redshift 
dependence, $\beta$. The values in this table correspond to the specific 
case of DES with $\ellmax=5000$. The other cases were explore yield very 
similar best-fit concentrations.
}
\vspace*{8pt}
\begin{tabular}{r c c c }
\hline
\hline
Simulation \ \ & \ \ $c_0/c_0^{\tt DMONLY}$ \ \  &\ $\alpha$ &\ $\beta$ \\ 
\hline
{\tt REF} & 1.13 & 0.20 & 0.83 \\
{\tt WML4} & 1.14 & 0.18 & 0.91 \\
{\tt NOSN} & 1.20 & 0.27 & 0.50 \\
{\tt NOZCOOL} & 1.16 & 0.16 & 0.91 \\
{\tt WDENS} & 0.96 & 0.20  & 1.39 \\
{\tt WML1V848} & 1.09 & 0.22 & 1.33 \\
{\tt DBLIMFV1618} & 0.89 & 0.24 & 1.23 \\
{\tt NOSN\_NOZCOOL} & 1.41 & 0.40 & 1.44 \\
{\tt AGN} & 0.34 & 0.79 & 2.06\\
\hline
\end{tabular}
\label{table:bfparams}
\end{table}
%%%%%%%%%%%%%%%%%%%%%%%%%%%%%%%%%%%%%%%%%%%%%%%%%%%%%%%%%%%%%%%%%%%%%%

Motivated by prior studies indicating that the largest effect of baryons 
on convergence spectra on relevant scales is a 
modification of halo structure \cite{Rudd:2007zx,Zentner:2007bn}, 
we pursue a mitigation strategy in which baryonic effects are entirely encapsulated 
into changes in the internal mass distributions within dark matter halos.  
It is certainly not true that the {\em only} effect of baryons is to 
alter halo structures. For example, the distribution of halo masses changes slightly  
(e.g., Ref.~\cite{Rudd:2007zx,stanek_etal09}), 
and baryonic effects extend beyond halo virial radii 
(e.g., Ref.~\cite{Rudd:2007zx,vanDaalen:2011xb}). Our goal 
is to determine the practical utility of such a model in analyses of 
forthcoming data.

We assume that the average mass distributions within dark matter halos 
can be described by the density profile of Ref.~\cite{Navarro:1997he} (NFW hereafter), 
\be
\label{eq:nfw}
\rho(r) \propto \frac{1}{(cr/\Rvir)(1+cr/\Rvir)^2}.
\ee
The parameter $c$ is the halo concentration and the density profile is 
normalized by our definition of a halo as a spherical object within which 
the mean density is $200$ times the mean density of the universe, $\rhovir = 200 \rhomean$.  
Therefore, halo mass and radius are related by $m=4\pi \rhovir \Rvir^3/3$, so that 
the profile can be normalized by $m = 4\pi\int_0^{\Rvir}\, \rho(r)\, r^2\, \dd r$ 
for a given mass and concentration. Halo concentrations predicted by dissipationless 
simulations of dark matter only have been studied extensively 
(recent examples are Refs.~\cite{neto_etal07,maccio_etal07,maccio_etal08,munoz-cuartas_etal11,prada_etal12}).  
The relationship between halo concentration, halo mass, and redshift in the 
OWLS DMONLY simulation can be adequately characterized by a 
power-law distribution \cite{duffy_etal08,duffy_etal10,vanDaalen:2011xb}\footnote{
The fit values come from Ref.~\cite{duffy_etal08} after applying a 
correction to change the pivot mass to $M_p$ at $z=0$. 
Ref.~\cite{duffy_etal08} explored a set of simulations that differ slightly 
from the OWLS simulations that we examine here. Ref.~\cite{duffy_etal10} and 
Ref.~\cite{vanDaalen:2011xb} 
confirm that the OWLS DMONLY simulation has a very similar concentration-mass 
relation, but do not provide detailed fits to the concentration-mass relation 
as a function of redshift. For the purposes of the present paper, 
this relation serves only to establish a baseline 
with respect to which we model the remaining baryonic 
simulations. Plausible alterations to the baseline model, given the 
results quoted in Refs.~\cite{duffy_etal08,duffy_etal10}, cause only 
minor quantitative changes to our subsequent results.
}, 
\be
\label{eq:cpl}
c(M,z)=c_0\, \left(\frac{M}{M_p}\right)^{-\alpha}\, (1+z)^{-\beta}, 
\ee
with $c_0=7.5$, $\alpha=0.08$, and $\beta=1$, in broad agreement with prior studies. 
The parameter $M_p$ is a pivot mass, which we take to be 
$M_p = 8 \times 10^{13}\, h^{-1}\mathrm{M}_{\odot}$. We choose the 
pivot mass to be close to the halo mass that is most well constrained 
by lensing spectra \cite{Zentner:2007bn}.

We describe modifications to halo structure through a modified 
concentration relation, following Refs.~\cite{Rudd:2007zx,Zentner:2007bn}.  
In particular, we allow the parameters $c_0$, $\alpha$, and $\beta$ in 
Eq.~(\ref{eq:cpl}) to vary in order to describe convergence power spectra 
within the baryonic OWLS simulations.  We determine the values that 
best capture the simulation results as follows.  
For each simulation, we produce a set of $\ntomo$ convergence power spectra. 
We fit the spectra by minimizing 
\begin{eqnarray}
\chi^2 & = & \sum_{\ell=\ell_{\mathrm{min}}}^{\ellmax}(2\ell + 1)\fsky \nonumber \\
       & \times &  \sum_{\mathrm{i,j,k,l}} \delta \Pktom{i}{j}(\ell)\, 
         \cov^{-1}[ \Pktom{i}{j}(\ell), \Pktom{k}{l}(\ell) ]\, \delta \Pktom{k}{l}(\ell),
\end{eqnarray}
where $\delta \Pktom{i}{j}(\ell)$ is the difference between the model and 
the simulation prediction, 
for the concentration parameters $c_0$, $\alpha$, and $\beta$ at {\em fixed} 
cosmology. This results in best-fit values for the concentration parameters, 
and residual differences between the best-fit modified concentration models 
and the predicted spectra from the baryonic simulations.  
In this manner, we assess the ability of the 
modified concentration model to describe the baryonic simulations.

The implementation of our model for the effect of modified 
concentrations on convergence power spectra is based upon the 
halo model for $P_{\delta}(k,z)$. The details of the halo 
model implementation are described in Ref.~\cite{Zentner:2007bn}.  
However, using the halo model to fit convergence power 
spectra from the OWLS simulations introduces a non-trivial 
complication.  On large scales, the halo model predictions 
for $P_{\delta}(k,z)$ are systematically offset from the OWLS 
simulation predictions because of the finite volumes of the simulations. 
To overcome this, we proceed as follows.  For each trial value 
of the halo concentration parameters, we compute a matter power 
spectrum offset $\Delta P_{\delta,\mathrm{HM}}(k,z)$ between the fiducial 
halo model with the standard values of $c_0$, $\alpha$, and 
$\beta$ and the halo model prediction with our trial values 
of the concentration parameters.  We then compute our 
trial matter power spectrum, which we use to predict convergence 
and compute $\chi^2$, by adding the offset defined by the 
halo model to the DMONLY prediction, 
$P_{\delta,\mathrm{trial}}(k,z) = P_{\delta,\mathrm{DMONLY}}(k,z) + \Delta P_{\delta,\mathrm{HM}}(k,z)$.  
In other words, we utilize the halo model to estimate a correction 
to be applied to the DMONLY matter power spectra.  In this manner, 
the spectra are not offset systematically with respect to each other 
on large scales. This strategy mimics what would likely have to be done 
in any analysis of real data; the predictions of dissipationless dark matter 
simulations would be established, but a correction would need to be applied to 
account for baryonic effects. This is the same strategy suggested in Ref.~\cite{Rudd:2007zx} 
and adopted by Ref.~\cite{semboloni_etal11}.

Figure~\ref{fig:fitband} summarizes the results we obtained by 
fitting concentrations to describe the baryonic OWLS simulations.  The 
figure shows the difference between the power spectra in the 
best-fit models with modified concentrations and the ``true'' 
baryonic simulations. This is analogous to Fig.~\ref{fig:resband}. 
As in Fig.~\ref{fig:resband}, Fig.~\ref{fig:fitband} shows only residuals 
of the auto-correlation for sources in the third DES photometric redshift bin, 
in the interest of clarity.  The most obvious feature of Fig.~\ref{fig:fitband} 
is that the fit residuals are nearly an order of magnitude smaller than the 
differences between the baryonic and DMONLY simulations.  The correlated 
structure of the fit residuals is also shown in Fig.~\ref{fig:fitband} by 
the two principal modes that account for the majority of the variance in the 
spectra among the baryonic OWLS simulations.  The two modes depicted in 
Fig.~\ref{fig:fitband} account for nearly $97\%$ of the variance among the 
spectra in the third redshift bin.

The residuals in Fig.~\ref{fig:fitband} represent the remaining 
systematic errors in the predictions of convergence power spectra 
after accounting for baryonic effects using a simple, phenomenological 
model. These systematic offsets will propagate into biases in the 
estimators of the cosmological parameters.  We explore this 
in detail in the next section.

The best-fit concentrations are given in Table~\ref{table:bfparams}. 
The concentrations of the dark matter halos in the OWLS simulations 
have been presented in \citet{duffy_etal10}. The concentrations 
we quote in Table~\ref{table:bfparams} exhibit two important deviations 
from the results quoted in Ref.~\cite{duffy_etal10}. First, our fit to 
the AGN simulation results in significantly lower concentrations. This could 
be because mass that would be within the DMONLY virial radii of these halos 
is moved outside of the virial radii in the AGN simulation and the concentrations 
are driven to low values to account for this effect. The OWLS 
collaboration has argued that the AGN simulation is the most realistic among 
their suite, so a genuine discrepancy in this case could have important 
implications for our method. Second, our fits generally yield 
a stronger mass dependence than the OWLS fits. 

A comparison of between our concentration results is not trivial for several 
reasons. Foremost among these, Ref.~\cite{duffy_etal10} quote the effective 
concentrations of dark matter while lensing is sensitive to {\em all} matter. 
The concentrations that we quote are the effective concentrations 
of all matter (baryonic as well as dark). Ref.~\cite{duffy_etal10} does 
provide non-parametric measures of total matter concentration; however 
these measures exhibit behavior different from the NFW fits and it is 
unclear how they correspond to the result of our fitting procedure. 
Additionally, the lensing signal on 
the scales that we explore is sensitive to mass redistribution on scales 
near halo virial radii and insensitive to halo profiles on significantly 
smaller scales. Therefore, it is not clear that our exercise should yield 
the same concentrations as those derived from fitting profiles directly 
because the two procedures are not equally sensitive to halo structure 
at all scales. Furthermore, we have assumed power law relationships 
between halo concentration, halo mass, and redshift. This can be justified 
by the fact that lensing is sensitive to a relatively narrow range halo 
masses and redshifts \cite{Zentner:2007bn}, so that the power-law indices 
that we recover may not correspond to those derived from a fit to simulation 
results over a wide range of masses and redshifts. As a results of these 
complicating factors, we reserve a more detailed comparison between our concentration 
parameters and those of the simulations for future work. 

%-----------------------------------
% bias
\begin{figure*}[t!]
\includegraphics[width=16cm]{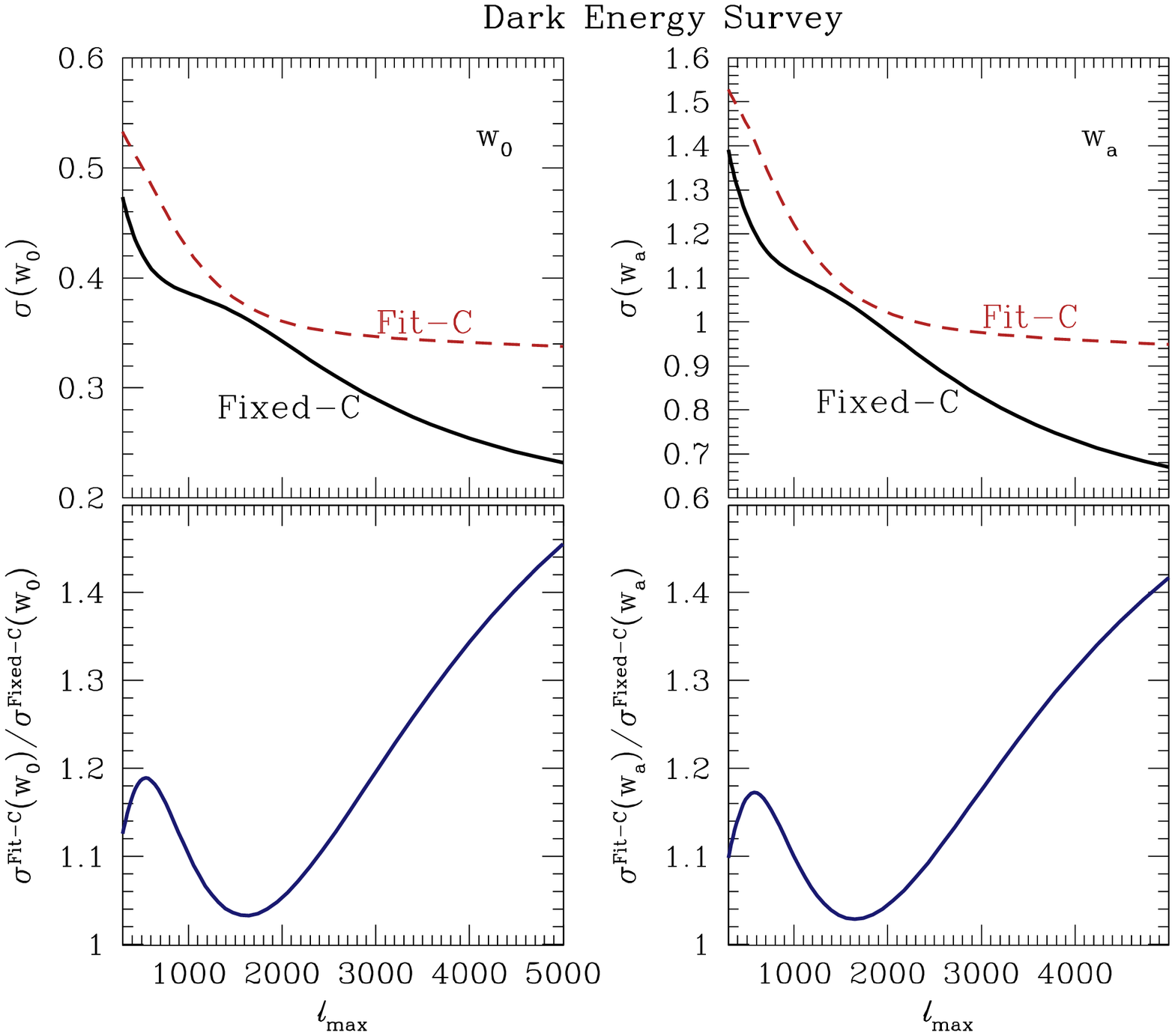}
\caption{ 
Constraints as a function of maximum multipole used 
to infer cosmological parameters.  The {\em left} panels 
show results for $\wzero$, while the {\em right} panels 
show results for $\wa$.  In the {\em top} panels, the 
{\em solid} lines show standard constraints assuming 
that the power spectrum is known perfectly.  The {\em dashed} 
lines show the constraints that may be realized if the 
halo concentration-mass relation must be fit simultaneously 
with cosmological parameters.  The {\em bottom} panels 
show the ratio of the constraints realized when 
concentrations must be fit, $\sigma^{\mathrm{Fit-C}}$ 
to the constraints realized 
in the standard scenario, $\sigma^{\mathrm{Fixed-C}}$.  
Constraints are significantly degraded beyond 
$\ellmax \sim 3000$ when the concentration-mass relation 
is permitted to vary.  Constraints on $\wzero$ and 
$\wa$ show the same qualitative features.  
}
\label{fig:condeg}
\end{figure*}
%------------------------------------

%--------------------------------------------------------------
\section{Cosmological Constraints and Residual Bias}
\label{section:results}

In this section, we project the effects of baryons in simulations onto 
cosmological parameters.  We begin by discussing our results in the 
context of the DES and conclude the section with a brief discussion 
of possible cosmological biases for stage IV experiments such as 
LSST or Euclid.

It is useful to consider the baseline 
constraints on the dark energy equation of state parameters.  
We consider two cases that will prove useful in the following.  
The first case corresponds to standard constraint projections 
on the dark energy equation of state parameters assuming that the 
nonlinear growth of structure is known perfectly.  In the context 
of our analysis, this means that the concentrations of halos are 
known perfectly and we refer to these constraints as ``Fixed-C'' 
constraints accordingly.  As we are exploring a mitigation strategy 
in which we fit for concentrations concurrently with cosmological 
parameters, it is necessary to assess the degradation in dark 
energy parameter constraints due to this additional freedom. We 
refer to constraints derived from an analysis in which concentrations 
are fit alongside cosmological parameters as ``Fit-C''.

%-----------------------------------------------------------------------
\subsection{Results for the Dark Energy Survey}
\label{sub:des}
%-----------------------------------------------------------------------

Our baseline constraints for DES, as well as the degradation in 
constraints when concentrations are fit alongside cosmological 
parameters, are shown as a function of the maximum 
multipole used in the analysis in Figure~\ref{fig:condeg}. 
Fig.~\ref{fig:condeg} contains four panels. The two panels 
on the left show results for $\wzero$, while the two 
panels on the right show results for $\wa$. The top panels 
show the marginalized constraints on the equation of state 
parameters as a function of the maximum multipole used 
in the analysis. The solid lines show the marginalized 
constraints in the standard Fixed-C case, while the 
dashed lines show constraints in the Fit-C case in which 
concentrations are permitted to vary. The lower panels 
show the ratio of the Fit-C constraints to the Fixed-C 
constraints at each multipole, giving the factor by which 
introduction of the additional nuisance parameters describing 
concentrations degrades the constraints.

It is clear that constraints are 
degraded if the concentration-mass relation of dark 
matter halos must be allowed to vary.  This degradation 
is mild ($\lesssim 20\%$) if the maximum multipole 
used in the cosmological analysis is $\ellmax \lesssim 3000$, 
and increases to $\gtrsim 40\%$ once scales to 
$\ellmax \approx 5000$ are included. 
The level of degradation depicted in Fig.~\ref{fig:condeg} is slightly larger 
than that estimated in Ref.~\cite{Zentner:2007bn} for a DES-like experiment.  
We find that this discrepancy is almost entirely due to the fact that the 
source redshift distribution used in the present study differs significantly 
from that assumed in Ref.~\cite{Zentner:2007bn}.  In particular, the redshift distribution 
that we assume concentrates source galaxies at significantly lower redshift, 
resulting in relatively lower lensing power compared to noise and reducing 
the lever arm to high redshift sources.

To determine the impact of this mitigation scheme on the dark energy program 
of a particular experiment, such as DES, the degradation in cosmological parameters 
caused by fitting for the concentrations of halos must be compared to 
the biases in these parameters that may be realized if baryonic effects are 
neglected.  We compute these biases by using the residuals between the 
DMONLY and baryonic simulations as the systematic offsets in Eq.~(\ref{eq:fisherbias}).  
An example of these offsets is depicted in Fig.~\ref{fig:resband}. 

%-----------------------------------
% bias
\begin{figure}[t!]
\includegraphics[width=9cm]{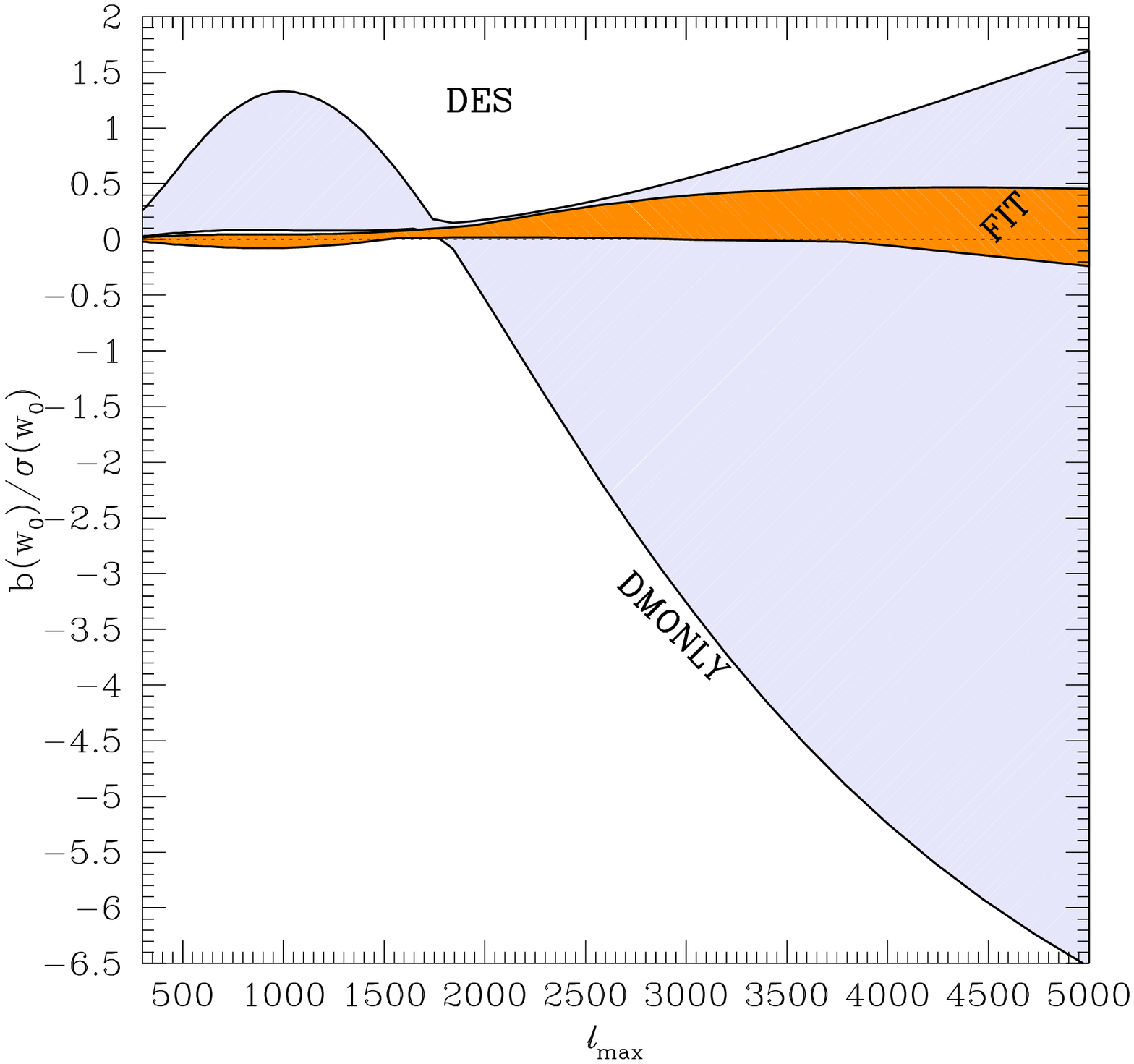}
\caption{ 
Biases induced in the dark energy equation of state parameter $\wzero$ as a function 
of the maximum multipole used to infer cosmological parameters.  The 
bias is shown in units of the statistical error on $w_0$ in order to make the 
relative importance of the systematic error induced by baryons explicit.  
The outer shaded (blue) band covers the range of biases spanned by computing the 
biases induced by analyzing all of the OWLS baryonic simulations without any 
model for baryonic effects.  The inner shaded (orange) band shows the range of biases 
induced by after taking the best-fit concentration model to describe baryonic effects 
in the OWLS simulations.  
}
\label{fig:biasbandw0}
\end{figure}
%------------------------------------

\begin{figure}[t!]
\includegraphics[width=9cm]{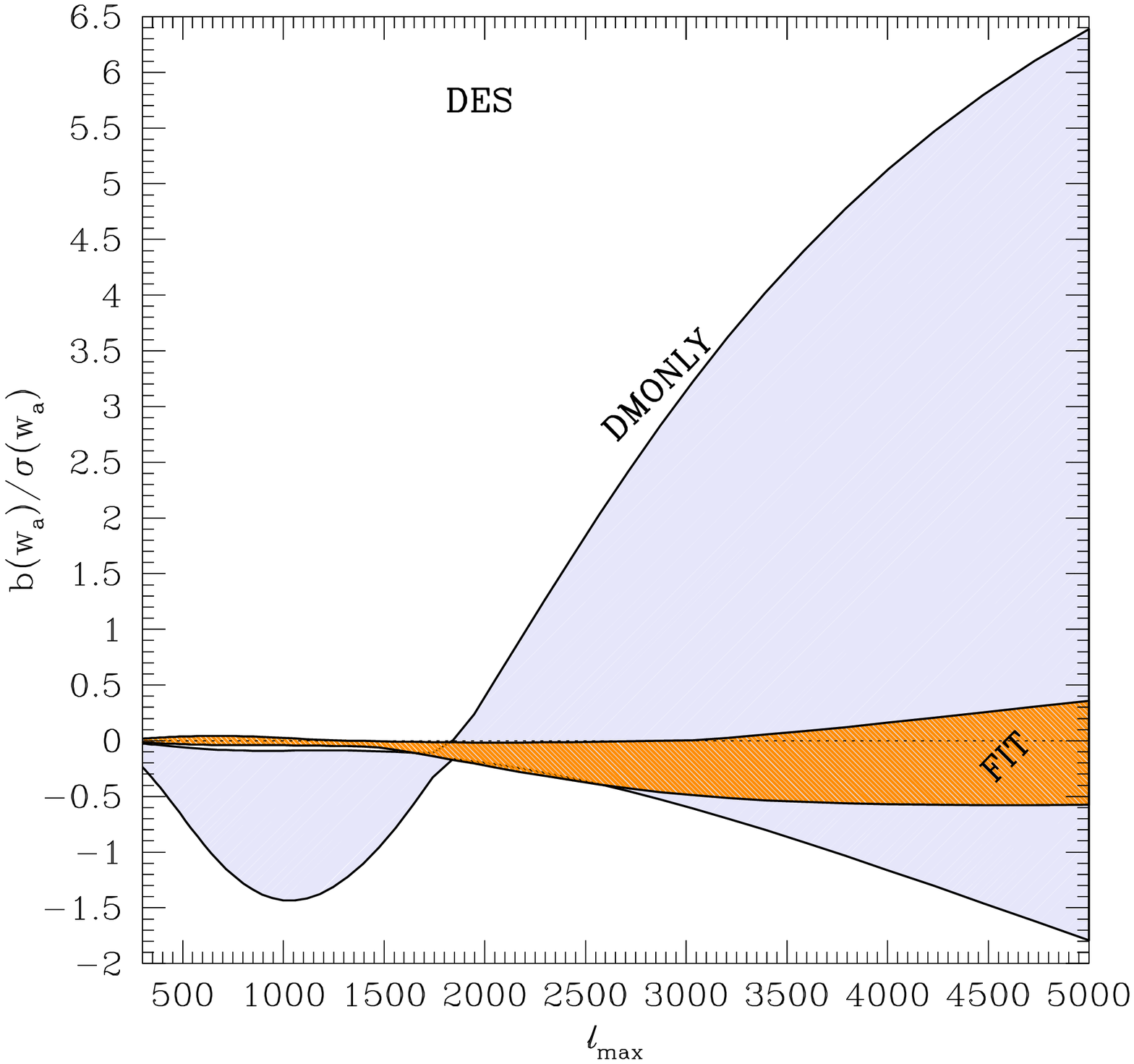}
\caption{ 
Biases induced in the dark energy equation of state parameter $\wa$ as a function 
of the maximum multipole used to infer cosmological parameters.  The 
bias is shown in units of the statistical error on $w_a$ in order to make the 
relative importance of the systematic error induced by baryons explicit.  
The outer shaded (blue) band covers the range of biases spanned by computing the 
biases induced by analyzing all of the OWLS baryonic simulations without any 
model for baryonic effects.  The inner shaded (orange) band shows the range of biases 
induced by after taking the best-fit concentration model to describe baryonic effects 
in the OWLS simulations.  
}
\label{fig:biasbandwa}
\end{figure}

The maximal biases that are induced by neglecting baryonic effects 
in our analysis of the OWLS simulations are shown as the outer (blue) bands in 
Fig.~\ref{fig:biasbandw0} (for $\wzero$) and Fig.~\ref{fig:biasbandwa} (for $\wa$).  
In particular, the {\em outer} ({\em blue}) bands delineate the extremal biases 
(maximum and minimum as the biases may be positive or negative) 
induced by analyzing any of the OWLS simulations without accounting 
for baryonic effects. The outer bands in Fig.~\ref{fig:biasbandw0} and 
Fig.~\ref{fig:biasbandwa} represent the envelope of the bias from all 
simulations, while for any individual simulation, the bias is a 
smooth function of maximum multipole.  The features in the bands in 
Fig.~\ref{fig:biasbandw0} and Fig.~\ref{fig:biasbandwa} arise when 
the particular simulation that gives rise to the extremal bias changes from 
one multipole to the next.

The biases induced by neglecting baryonic effects in Fig.~\ref{fig:biasbandw0} and 
Fig.~\ref{fig:biasbandwa} are as large as $\sim 3\sigma$ for 
$\wzero$ and $\wa$ if the analysis include multipoles out to 
$\ellmax \approx 3000$. Including multipoles out to $\ellmax \approx 5000$ 
drives the maximal potential bias to $\sim 6\sigma$. These bias levels 
have a clear significance for the effort to understand dark energy. 
However, we remind the reader that we 
have used the Fisher matrix approximation to 
estimate the biases on cosmological parameters. A necessary caveat 
to our results is that Eq.~(\ref{eq:fisherbias}) is the lowest-order 
approximation to the bias in the limit of small parameter biases and 
may not provide an accurate bias estimate for large biases. 
Nevertheless, the statement that the biases are significant ($\gtrsim 1\sigma$) 
in this case is robust.

%-----------------------------------
% bias
%------------------------------------

The inner (orange) bands in Fig.~\ref{fig:biasbandw0} and Fig.~\ref{fig:biasbandwa} 
delineate the extremal range of dark energy equation of state parameter biases 
realized after fitting for the halo concentration-mass relation in the baryonic 
simulations and using the fit to correct the power spectra as described in 
\S~\ref{section:fit}.  Specifically, we compute these biases by utilizing the 
residuals between the {\em corrected} DMONLY spectra and the baryonic simulations, 
an example of which is shown in Fig.~\ref{fig:fitband}, in Eq.~(\ref{eq:fisherbias}). 
To make Fig.~\ref{fig:biasbandw0} and Fig.~\ref{fig:biasbandwa} show fair comparisons 
of the biases, the biases for the fit-corrected cases are shown in units of the 
statistical error in the case of fixed concentrations (``Fixed-C''). Showing these biases in units 
of the statistical error in the case of varying concentrations (``Fit-C'') would 
reduce their magnitudes in Fig.~\ref{fig:biasbandw0} and Fig.~\ref{fig:biasbandwa}.

Fitting for concentrations clearly leads to dramatic reductions in parameter biases.  
Indeed, the biases are typically less than $\sim 10\%$ of the statistical error at low multipoles 
and never exceed $\sim 50\%$ ($\sim 60\%$) of the statistical error in $\wzero$ ($\wa$) 
for $\ellmax \lesssim 5000$. This suggests that the mitigation strategy of 
fitting for a halo concentration relation alongside cosmological parameters will 
result in a dark energy error budget that is preferable to neglecting baryonic 
effects. At $\ellmax \sim 3000$, fitting for concentration increases the statistical 
error in $\wzero$, for example, by $\sim 20\%$ compared to the ideal case 
(Fig.~\ref{fig:condeg}), and reduces the systematic error to $\sim 40\%$ of the 
statistical error (maximum). Taking a simple and conservative approach of adding 
these two contributions, the resulting error on $\wzero$ increases to 
$\sim 160\%$ of the constraint in the ideal case. This is to be compared 
to a potential systematic error in the case where 
no mitigation for baryonic processes is undertaken of as much as 
$\sim 350\%$ of the statistical error.

%-----------------------------------
\subsection{Stage IV Experiments}

\begin{figure*}[t!]
\includegraphics[width=16cm]{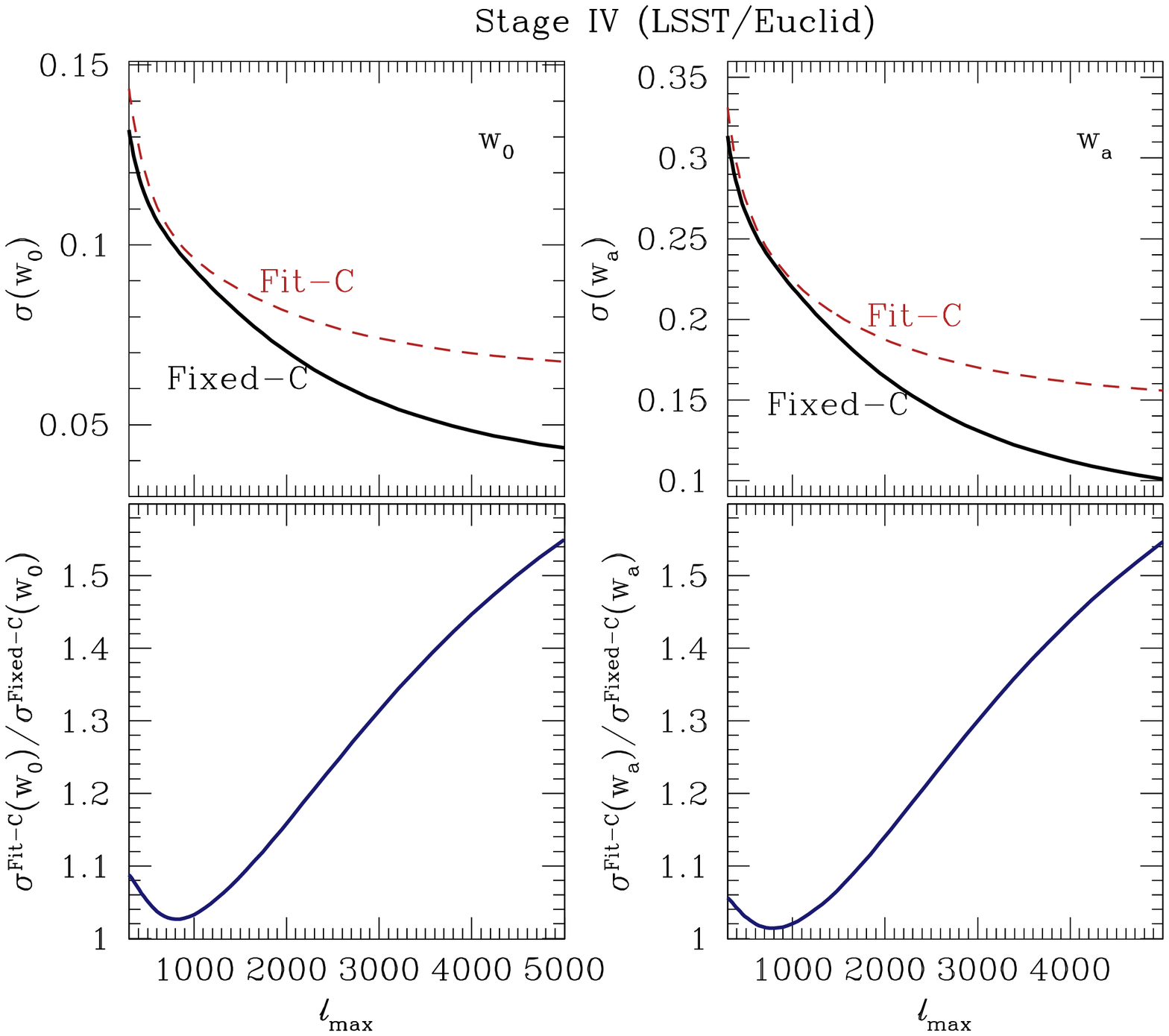}
\caption{ 
Constraints as a function of maximum multipole used 
to infer cosmological parameters.  The panels and 
lines are as in Fig.~\ref{fig:condeg}, but pertain 
to a Stage IV dark energy experiment such as the 
Large Synoptic Survey Telescope or the European Space 
Agency's Euclid.  
}
\label{fig:condeglsst}
\end{figure*}

Having discussed the utility of the proposed mitigation scheme of 
\citet{Zentner:2007bn} for DES, we briefly describe analogous results 
for forthcoming, Stage IV dark energy experiments such as LSST and Euclid.  
Figure~\ref{fig:condeglsst} shows cosmological constraints for a Stage IV 
dark energy experiment, including the degradation in those constraints 
incurred by fitting for the concentrations of halos concurrently with 
the cosmological parameters.  Fig.~\ref{fig:condeglsst} exhibits two notable 
features compared to the analogous results for DES (Fig.~\ref{fig:condeg}).  
First, the constraints from Stage IV experiments are significantly more 
restrictive, though this is an expected result (e.g., Ref.~\cite{Albrecht:2006}).  
Second, Stage IV experiments suffer from slightly greater degradation 
in dark energy parameter constraints when fitting for halo structure 
along with cosmological parameters.  In particular, the degradation in 
dark energy parameters reaches $\approx 30\%$ at a maximum multipole 
of $\ellmax \approx 3000$ and $\approx 55\%$ at $\ellmax \approx 5000$.  

%-----------------------------------
% constraints for lsst
%------------------------------------

\begin{figure}[t!]
\includegraphics[width=9cm]{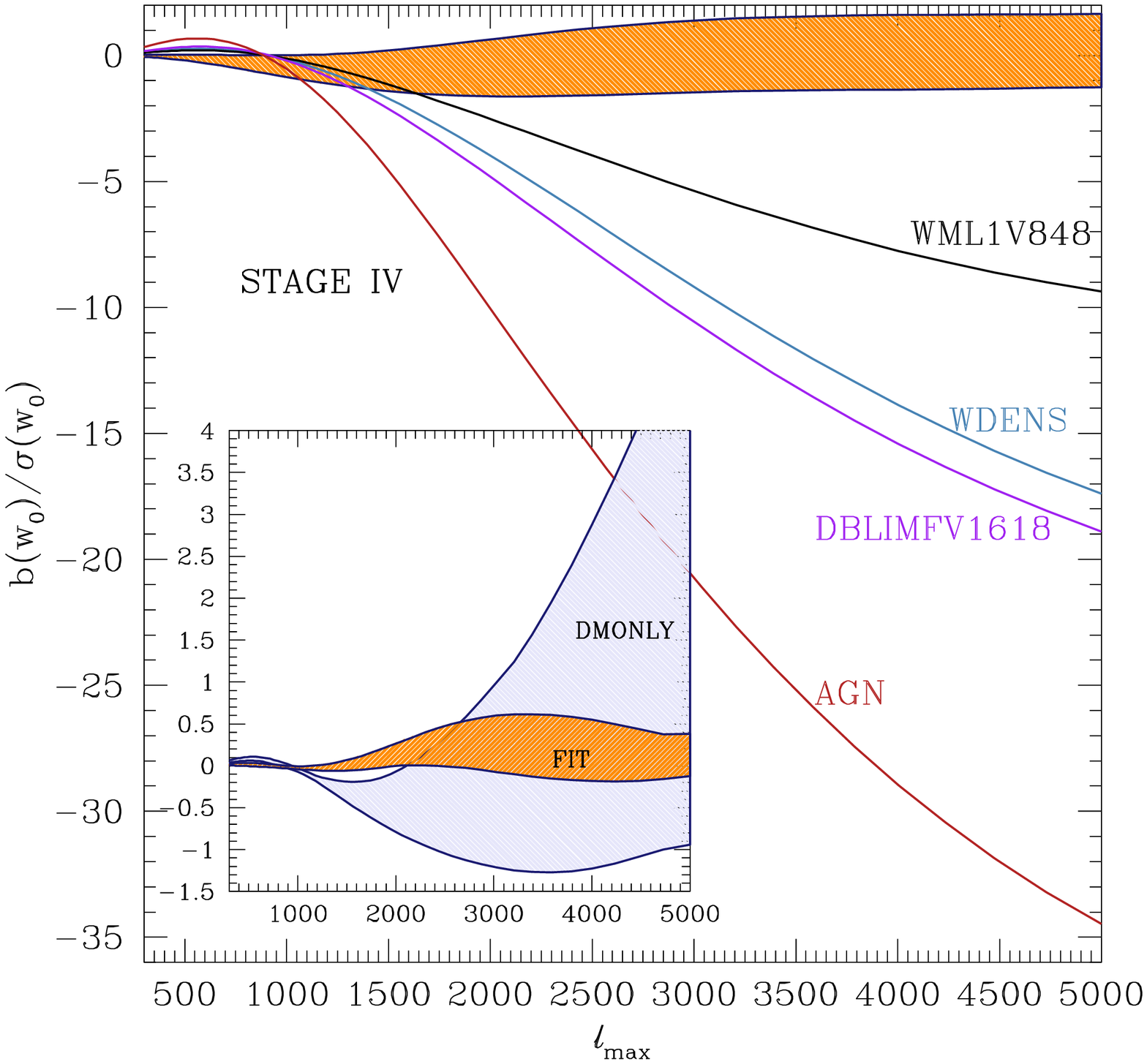}
\caption{
Biases induced in the dark energy equation of state parameter $\wzero$ from the 
analysis of a STAGE IV dark energy experiment as a function 
of the maximum multipole used to infer cosmological parameters. 
For STAGE IV experiments, four of the OWLS simulations lead to 
biases significantly larger than the others and it is useful to 
emphasize this. In the main panel, 
the biases that result from analyzing those four simulations 
without any model to account for baryonic effects. These biases 
are clearly very large. Each line is labeled by the name of the 
corresponding OWLS simulation in the panel. 
The shaded (orange) band, shows the 
range of biases that result after taking the best-fit concentration 
model to describe baryonic effects to analyze these same four simulations. 
The biases here are significantly reduced, but remain non-negligible ($\sim 1\sigma$). 
The inset panel shows results for the remaining five OWLS simulations. 
In this case, the inner (orange) and outer (blue) shaded bands are the 
same as in Fig.~\ref{fig:biasbandw0} for DES. In each of these cases, 
the mitigation procedure renders biases in the dark energy equation of 
state parameter $\wzero$ smaller than the statistical error.
}
\label{fig:bbw0lsst}
\end{figure}

\begin{figure}[t!]
\includegraphics[width=9cm]{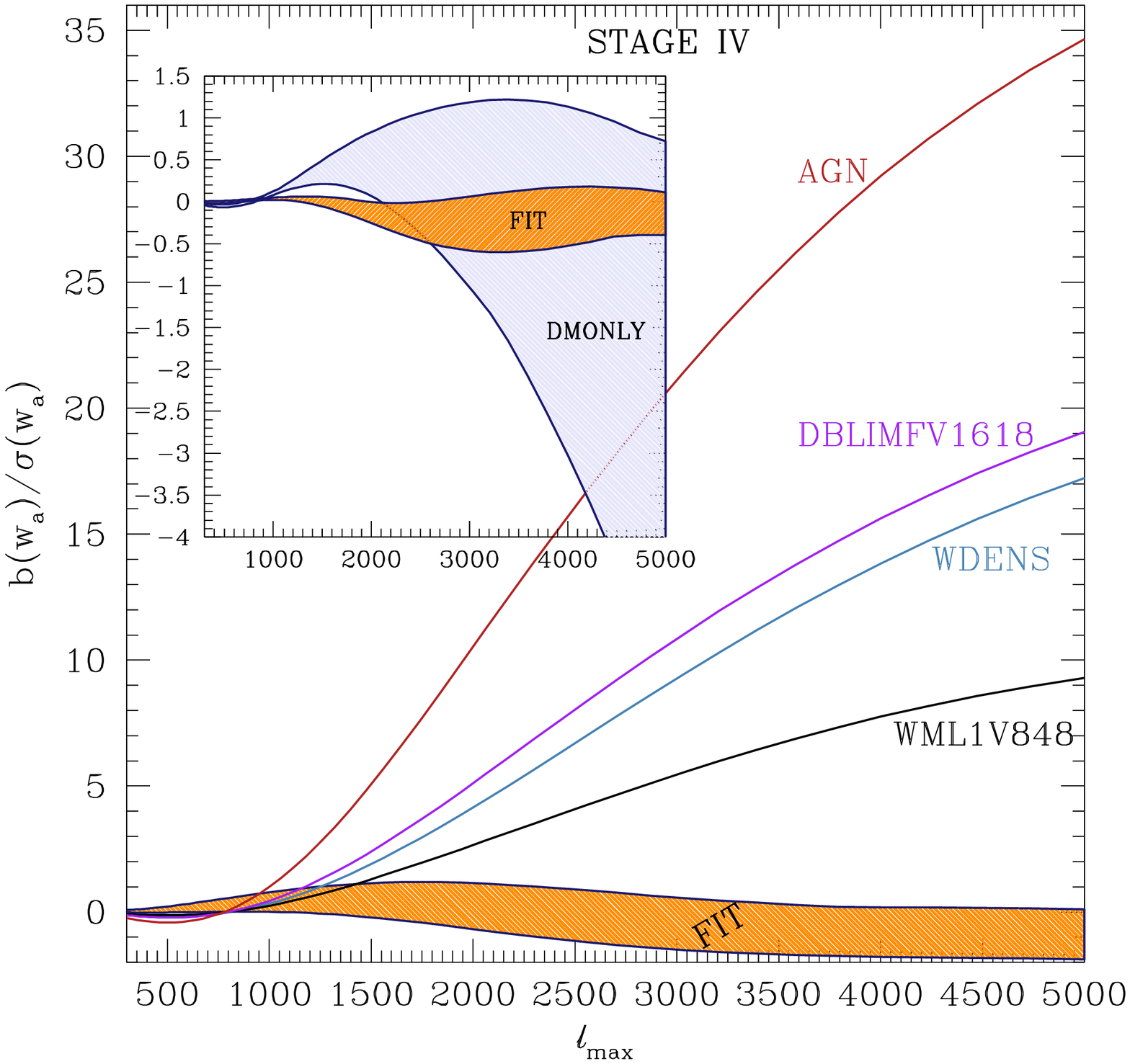}
\caption{ 
Same as Fig.~\ref{fig:bbw0lsst}, but for the 
bias on $\wa$ from a Stage IV experiment. 
Notice that the vertical axis is asymmetric about zero. 
}
\label{fig:bbwalsst}
\end{figure}

The biases resulting from the analysis of STAGE IV experiments are 
shown in Figure~\ref{fig:bbw0lsst} ($\wzero$) and 
Fig.~\ref{fig:bbwalsst} ($\wa$). Notice that we present the 
results in a slightly different way for STAGE IV experiments. 
In particular, four of the OWLS simulations 
(``AGN'', ``DBLIMFV1618'', ``WDENS'', and ``WML1V848'') 
give significantly larger biases than any of the other 
five simulations. Therefore, we show the biases for 
these individually in the main panels of Fig.~\ref{fig:bbw0lsst} 
and Fig.~\ref{fig:bbwalsst}. We show results for the 
remaining five simulations in the inset panels. 
In all cases, the value of fitting concentrations to 
mitigate for baryonic effects in cosmological parameter 
analyses is apparent. However, notice that the residual 
biases in the worst cases can remain significant compared 
to the ideal statistical error even after fitting for 
the concentration-mass relation. This indicates that 
a more accurate mitigation scheme will be needed 
in order to reduce the theoretical systematic associated 
with baryonic physics to the level of the statistical 
errors expected of STAGE IV dark energy experiments.

%-----------------------------------
% bias
%------------------------------------

%-----------------------------------
% bias
%------------------------------------

%---------------------------------------------------------------
\section{Conclusions}
\label{section:conclusions}

We have explored the viability of a strategy to mitigate the influence 
of baryonic effects on dark energy constraints from cosmological weak 
lensing.  The strategy entails fitting lensing data for both cosmological 
parameters and the concentration-mass 
relation of all matter in halos simultaneously.  
We assessed this scheme by using it to analyze power spectra 
predicted by a suite of cosmological simulations as though they 
were genuine data. Specifically, we computed the 
resultant systematic and statistical errors on dark energy 
parameters that would be incurred by fitting such data according 
to this strategy.

We find that introducing additional parameter freedom to describe 
the concentration-mass relations of halos reduces the systematic 
errors on dark energy parameters to marginally-acceptable levels 
in all cases. For a DES-like analysis exploiting all multipoles $\ell \le 3000\ (5000)$, 
fitting for concentrations increases statistical 
errors by $\lesssim 25\% \ (45\%)$ (Fig.~\ref{fig:condeg}), 
but reduces the potential systematic error by a factor 
of as much as $\sim 7$, to $\lesssim 0.3\sigma\ (0.5\sigma)$, in the worst case scenario
(Fig.~\ref{fig:biasbandw0} and Fig.~\ref{fig:biasbandwa}). 
The reduction in systematic error outweighs the increase in statistical 
error suggesting that this mitigation scheme may be a viable option for 
analyzing data of the quality expected from DES.

For Stage IV experiments, such as the surveys to be 
undertaken by the LSST or Euclid, the conclusion is 
somewhat less straightforward. What is clear is that 
some mitigation strategy for baryonic effects is 
necessary. Systematic errors on $\wzero$ and $\wa$ 
incurred by fitting the OWLS simulations with no 
model for baryonic processes can be as large as 
several tens of the statistical error if all scales 
$\ell \lesssim 5000$ are included in the analysis 
(Fig.~\ref{fig:bbw0lsst} and Fig.~\ref{fig:bbwalsst}). 
Of course, as we mentioned in the introductory section, 
such large biases are not likely to be realized as the 
result of any analysis. Rather, it is likely that the 
analysis team will not find acceptable fits to the 
observables according to a specific fit criterion. 
Nevertheless, it is apparent that a model for possible 
baryonic effects will be necessary in order to 
extract cosmological parameters reliably.

For Stage IV experiments, the strategy of fitting 
the concentrations of halos in order to militate 
against large biases in the inferred cosmological 
parameters, particularly the dark energy equation of 
state parameters, is relatively less effective. 
One complicating factor is that the different OWLS simulations 
lead to more disparate conclusions in this case. 
In the worst case, that of the OWLS ``AGN'' simulation, 
the residual biases after fitting for concentrations 
are $\sim 1.6\sigma$, assuming all scales to $\ellmax = 5000$ 
are included. It is necessary to restrict consideration to 
multipoles $\ellmax \lesssim 1100$ in order to reduce this 
bias to $\sim 1\sigma$. However, for six of the nine 
OWLS simulations that we have analyzed, the residual 
bias including all scales to $\ellmax = 5000$ 
is $\lesssim 0.5\sigma$. The concomitant 
cost of the additional parameters for the 
{\em statistical} errors is $\lesssim 55\%$. 
Fitting for an effective halo concentration-mass 
relation does reduce biases in the dark energy 
equation of state parameters; however, in the most 
extreme cases that we have analyzed, these biases 
remain significant compared to the 
statistical errors expected from Stage IV experiments.

As pointed out by \citet{Zentner:2007bn}, fitting for an 
effective concentration-mass relation also yields information 
that may help to constrain galaxy formation models. In this case, 
the procedure gives constraints on the parameters of the concentration-mass 
relation at no additional cost. In the case of DES (Stage IV), the 
best-constrained halos have masses 
$M \sim 8 \times 10^{13}\, h^{-1}\mathrm{M}_{\odot}$ ($M \sim 6 \times 10^{13}\, h^{-1}\mathrm{M}_{\odot}$) 
at redshift $z \sim 0.23$ ($z \sim 0.31$) and constraints on the average 
concentrations of such halos are $\sigma_{c}/c \sim 0.06$ ($\sigma_{c}/c \sim 0.03$). 
Such constraints may prove useful in understanding the formation 
histories of galaxies and galaxy clusters.

A handful of other recent studies have investigated methods for 
marginalizing over uncertainty in power spectra in deriving 
cosmological constraints from weak lensing 
\cite{bernstein09,kitching_taylor11,hearin_etal12,semboloni_etal11,semboloni_etal12}. 
Refs.~\cite{bernstein09,kitching_taylor11,hearin_etal12} explore 
significantly more general parameterizations. However, they 
all reach conclusions that are broadly consistent with ours in 
that each finds self-calibration of uncertainty in the 
nonlinear matter power spectrum a promising approach. 
This broad agreement among different approaches likely 
stems from the well-known fact that cosmological information can 
be extracted from lensing data based only upon geometrical 
considerations \cite{jain_taylor03,zhang_etal05}. Refs.~\cite{semboloni_etal11,semboloni_etal12} 
are most similar to ours. These authors explore a halo model-based 
mitigation scheme in which gas and stars are modeled separately 
from dark matter, similar to the model proposed by \citet{Rudd:2007zx}. 
\citet{semboloni_etal11} find that their simple model can 
significantly reduce dark energy parameter biases for near term 
surveys, but that improvement may be necessary in order to address 
Stage IV dark energy experiments, a result in broad agreement with 
ours. However, \citet{semboloni_etal11} did not use their methods 
to model an {\em independent} set of simulations, nor did they 
address the statistical cost of marginalizing over additional 
parameters in their model.

In summation, our results suggest concentration fitting as a useful and viable 
strategy with which to analyze cosmological weak lensing power spectra from 
DES in order to extract constraints on the dark energy equation of state. 
Based on our analyses, Stage IV experiments may remain vulnerable to 
significant biases in the inferred values of the dark energy parameters 
even after militating against baryonic effects with concentration fitting. 
At minimum, estimates for systematic errors such as those presented here 
should be a component of the error budgets of such experiments.

Future work may be able to improve this situation. For one, simulations 
such as the OWLS simulations make predictions for the properties of galaxies. 
It may be possible to compare the properties of galaxies in order to 
determine which simulations are more likely to represent the observed universe, 
and use this information to place priors on additional parameters in 
mitigation schemes (the concentration parameters in our case, see Ref.~\cite{Zentner:2007bn}). 
The OWLS collaboration has shown that the ``AGN'' simulation describes the observed 
properties of galaxies most successfully \cite{mccarthy_etal10,vanDaalen:2011xb}, while 
our analysis of the ``AGN'' simulation for Stage IV experiments leaves a non-negligible 
residual bias. An important and necessary aspect of future efforts to address these 
issues with simulations will be to develop lensing predictions from baryonic simulations 
in larger computational volumes. On another front, 
it may be possible to develop more sophisticated models 
for the influence of baryons on lensing power spectra that can minimize biases 
in inferred cosmological parameters without a significant cost in statistical 
errors. As the cosmological community learns from Stage III experiments such as 
DES and prepares for the Stage IV experiments of the coming decade, such efforts 
should be a high priority in order to maximize the scientific yields of the 
next generation of dark energy experiments.

\begin{acknowledgments}
This work grew out of a working group meeting hosted by the Aspen Center for Physics. 
As such, this material is based upon work supported in part by the National Science 
Foundation under Grant PHY-1066293 and the hospitality of the Aspen Center for Physics. 
We are grateful to Marcel van Daalen, Joop Schaye, and the other members of the 
OWLS collaboration for making their simulation power spectra available.  We thank 
Henk Hoekstra, Dragan Huterer, Jeffrey Newman, Bob Sakamano, 
Joop Schaye, and Risa Wechsler for helpful discussions and comments on 
early drafts of this manuscript. 
ARZ and APH were funded by the Pittsburgh Particle physics, 
Astrophysics, and Cosmology Center (PITT PACC) at the 
University of Pittsburgh and by the National Science Foundation through 
grant AST 0806367. APH is also supported by the U.S. Department of Energy 
under contract No. DE-AC02-07CH11359. 
ES acknowledges support from the Netherlands Organisation 
for Scientific Research (NWO) grant number 639.042.814 and from 
the European Research Council under the EC FP7 grant number 279396. 
ES also acknowledges support from the Leids Kerkhoven-Bosscha foundation. 
SD was supported by the U. S. Department of Energy, including
grant DEFG02-95ER40896, and by the National Science
Foundation under Grant AST-090872. 
The research of TE and EK was funded in part by 
NSF grant AST 0908027 and U. S. Department of Energy 
grant DE-FG02-95ER40893. Support for this work was provided, 
in part, through the Scientific Discovery through Advanced 
Computing (SciDAC) program funded by the U. S. Department of Energy, 
Office of Science, Advanced Scientific Computing Research and High 
Energy Physics.
\end{acknowledgments}

\bibliography{v3}

\end{document}